\documentclass[fleqn,usenatbib]{mnras}
\pdfoutput=1
\usepackage{newtxtext,newtxmath}
\usepackage{tabularx} 
\usepackage{float}
\newcolumntype{Y}{>{\centering\arraybackslash}X}

\usepackage[T1]{fontenc}

\DeclareRobustCommand{\VAN}[3]{#2}
\let\VANthebibliography\thebibliography
\def\thebibliography{\DeclareRobustCommand{\VAN}[3]{##3}\VANthebibliography}

\usepackage{graphicx}	
\usepackage{amsmath}	

\usepackage{caption}
\usepackage{float}
\usepackage{pgfplots}
\usepackage{amsmath,bm}

\title[High-resolution retrievals of WASP-121b]{High-resolution atmospheric retrievals of WASP-121b transmission spectroscopy with ESPRESSO: Consistent relative abundance constraints across multiple epochs and instruments}

\author[C. Maguire et al.]{
Cathal Maguire$^{1}$\thanks{E-mail:  \href{mailto:maguic10@tcd.ie}{maguic10@tcd.ie}},
Neale P. Gibson$^{1}$,
Stevanus K. Nugroho$^{2}$, 
Swaetha Ramkumar$^{1}$, 
Mark Fortune$^{1}$,\newauthor
Stephanie R. Merritt$^{3}$,
and Ernst de Mooij$^{3}$
\vspace{-0.25cm} \\ \\ 
$^{1}$School of Physics, Trinity College Dublin, University of Dublin, Dublin 2, Ireland
\\
$^{2}$Astrobiology Center, NINS, 2-21-1 Osawa, Mitaka, Tokyo 181-8588, Japan\\
$^{3}$Astrophysics Research Centre, School of Mathematics and Physics, Queen’s University Belfast, Belfast BT7 1NN, UK
}

\date{Accepted 2022 November 16. Received 2022 November 15: in original form 2022 August 12}
\pubyear{2022}

\begin{document}
\label{firstpage}
\pagerange{\pageref{firstpage}--\pageref{lastpage}}
\maketitle

\begin{abstract}
Recent progress in high-resolution transmission spectroscopy has offered new avenues in which to characterise the atmospheres of transiting exoplanets. High-resolution cross-correlation spectroscopy allows for the unambiguous detection of molecules/atoms. It has also been used to map both atmospheric dynamics and longitudinal variations in the abundance of species across the morning and evening limbs. We present multiple VLT/ESPRESSO observations of the ultra-hot Jupiter WASP-121b, from which we constrain relative abundances of various neutral metals consistently across all observations, whilst accounting for the distortion of the exoplanet's signal caused by traditional data processing techniques. We also constrain planetary orbital velocities and $T$-$P$ profiles. We compare our abundance constraints with previous constraints using VLT/UVES transmission spectroscopy of WASP-121b, and find our results to be consistent between observations, and also in agreement with stellar values for species previously detected in the atmosphere of WASP-121b. Our retrieval framework can also be used to identify potential exospheric species, resulting in extended absorption features beyond the transit equivalent Roche limit of WASP-121b ($R_{\rm eqRL}$ $\sim$ 1.3 $R_{\rm p}$). H$\alpha$, Fe II, and Ca II were found to extend to high altitudes ($1.54\pm0.04$ $R_{\rm p}$, $1.17\pm0.01$ $R_{\rm p}$, and $2.52\pm0.34$ $R_{\rm p}$, respectively), which are broadly consistent with literature values. The consistency of our constraints across multiple high-resolution observations is a strong validation of our model filtering and retrieval framework, as well as the stability of the atmosphere over the timescales of months/years, and could allow for planet formation processes to be inferred from future ground-based observations of exoplanetary atmospheres.

\end{abstract}




\section{Introduction}
High-resolution transmission and emission spectroscopy have proven to be valuable techniques in the effort to characterise the atmospheres of exoplanets, in particular the ultra-hot Jupiters (UHJs; $T_{\rm eq} > 2,200$ K), owing to their extreme proximity to their host stars, such that they undergo immense stellar irradiation. Under such extreme conditions, the constituent molecules/atoms experience thermal dissociation and ionisation \citep{Parmentier_2018}, allowing their absorption/emission signatures to become more readily accessible, particularly at optical wavelengths.\newline \newline \indent WASP-121b, is an ultra-hot Jupiter ($R_{\rm p}$ = 1.753$\pm$0.036~$R_{\rm Jup}$), orbiting an F6V type star, with an orbital period of $P\sim1.27$~days and an equilibrium temperature of 2358$\pm$52 K \citep{Delrez_2016, Bourrier_2020}, on a highly misaligned orbit close to its Roche limit. It has been the subject of extensive observational campaigns, in transmission \citep{Evans_2016,Evans_2018, Salz_2019, Sing_2019,Wilson_2021}, emission \citep{Evans_2017,Mikal_Evans_2019,Kovacs_2019,Mikal_Evans_2020}, as well as with phase curve measurements \citep{Bourrier_2020b,Daylan_2021,Mikal_Evans_2022}, in an effort to characterise its atmosphere.\newline \newline \indent In recent years, high-resolution (R $\ge 25,000$) cross-correlation spectroscopy (HRCCS) \citep{Snellen_2010,Brogi_2012,Nugroho_2017,Hoeijmakers_2018,Ehrenreich_2020} has proven to be a very efficient method in unambiguously identifying chemical species and characterising the atmospheres of HJs/UHJs. This technique uses the large Doppler shift of the orbiting planet's atmospheric transition lines relative to its host star, with a significant shift between subsequent exposures, such that the planetary signal can then be detected through cross-correlation with a Doppler shifted atmospheric model. HRCCS has also been developed upon to infer atmospheric dynamics \citep{Snellen_2010,Brogi_2012,Seidel_2020,Wardenier_2021}, as well as map longitudinal variations in the abundance profiles of chemical species \citep{Ehrenreich_2020, Kesseli_2021, Kesseli_2022, Prinoth_2022}. It has also been shown that for some UHJs, including WASP-121b, the atmospheric regions probed by transmission spectroscopy studies are different before and after mid-transit, due to the large rotation angles of these planets relative to the ``opening angle'' of their day-to-night terminator regions \citep{Wardenier_2021b}. \newline \newline \indent
The readily accessible absorption signatures of these extreme objects, particularly those of neutral metals which have generally avoided condensation at such high temperatures, have allowed a plethora of species to be identified in the atmosphere of WASP-121b using this method, among others, alongside traditional low-resolution studies. Such species include neutral atomic species; Fe I, Mg I, Cr I, V I, Ca I, Na I, H I, Ni I, V I, K I, and Li I \citep{Gibson_2020, Hoeijmakers_2020, Borsa_2020, Cabot_2020} and ionic species; Fe II, Mg II, and Ca II \citep{Sing_2019,Ben_Yami_2020,Borsa_2020,Merritt_2021}. \cite{Lothringer_2021} recently showed that the abundance of these metals, or refractories, relative to that of volatile species such as H$_2$O and CO, in the form of a ``refractory-to-volatile'' ratio, could be used to constrain the formation and migration pathway of HJ/UHJs. \newline \newline \indent \cite{Evans_2016} found additional absorption near 1.2~{\textmu}m in the atmosphere of WASP-121b, tentatively attributed to a combination of strong optical absorbers such as TiO and VO, which aimed to explain the presence of a thermal inversion layer, indicated by the dayside emission of H$_2$O observed with the Hubble Space Telescope (HST) WFC3 \citep{Evans_2017}. This additional absorption attributed to VO could not be confirmed, however, with a joint analysis of this data set and follow-up secondary eclipse observations with the same instrument \citep{Mikal_Evans_2020}. \cite{Evans_2018} also found no evidence for TiO absorption features in the transmission spectrum of WASP-121b, as well as a strong NUV slope attributed to the SH molecule. The former of which was interpreted as the condensation of Ti-bearing species at higher temperatures compared to V-bearing species. Follow up observations using HRCCS also reported non-detections of TiO, VO, and SH \citep{Merritt_2020,Hoeijmakers_2020}, with \cite{Hoeijmakers_2020} postulating that absorption by atomic metals is sufficient to explain the NUV slope detected by \cite{Evans_2018}.\newline \newline \indent Exospheric Fe II and Mg II, extending beyond the Roche lobe of WASP-121b, has also been detected \citep{Sing_2019},  with the presence of Fe II confirmed at high-resolution \citep{Ben_Yami_2020, Borsa_2020}. \citet{Borsa_2020} also found evidence of H$\alpha$ and Ca II absorption at high altitudes with ESPRESSO, indicative of an evaporating upper atmosphere (see also \citealt{Cabot_2020, Merritt_2021}). \newline \newline \indent A recent study by \cite{Mikal_Evans_2022} revealed a nightside temperature below the condensation temperatures of neutral Fe, Ca, Mg, and V. However, these species have previously been detected in the gas-phase along the day-night terminator, suggesting that vertical mixing must be operating efficiently in the atmosphere of WASP-121b to prevent day-night cold trapping processes. \newline \newline \indent However, despite its success, HRCCS is not sensitive to absolute line strengths, thus the cross-correlation signals of different observations, as well as different atmospheric models, are difficult to compare in a statistically principled manner. These limitations arise partly due to the degeneracies introduced by the loss of the planet’s continuum as a result of the standard data processing techniques at high-resolution \citep{Birkby_2018}, which differ from those typically performed at low-resolution. In order to overcome these limitations, whilst utilising the power of HRCCS, \cite{Brogi_Line_2019} first developed a statistically robust atmospheric retrieval framework, in order to constrain chemical abundances and $T$-$P$ profiles from high-resolution emission spectroscopy observations, by ``mapping'' the cross-correlation value of a given model to a likelihood value (see also \citealt{Gandhi_2019}). \cite{Gibson_2020} then developed upon this approach, accounting for time- and wavelength-dependent uncertainties when computing the likelihood, whilst also applying it to transmission spectroscopy observations. \cite{Nugroho_2020} further applied this technique to emission spectroscopy observations of KELT-20b/MASCARA-2b. \newline \newline \indent However, in any retrieval framework, it is critical that the underlying forward model is an accurate representation of the (noiseless) data, after the aforementioned preprocessing steps \citep{Brogi_Line_2019, Pelletier_2021}. This was the remaining limitation in the approach presented in \cite{Gibson_2020}, which was later addressed with a novel model filtering technique \citep{Gibson_2022}. This model filtering technique mimics the effects of detrending algorithms, such as Principle Component Analysis (PCA) or \textsc{SysRem} \citep{Tamuz_2005,Birkby_2013}, which are typically used to remove (quasi-)stationary stellar and telluric lines from the data. These algorithms also act to inadvertently distort the exoplanetary lines \citep{Birkby_2017,Brogi_Line_2019}, and thus it is necessary to apply the same preprocessing steps to our forward model.\newline \newline \indent Developing robust techniques to constrain relative elemental abundances from high-resolution ground-based observations \citep{Gandhi_2019,Line_2021,Pelletier_2021,Gibson_2022}, such as carbon-to-oxygen (C/O) or refractory-to-volatile ratios, allows us to infer their formation and evolution mechanisms \citep{Oberg_2011,Madhusudhan_2014,Mordasini_2016,Lothringer_2021}, which has been a long-standing goal of exoplanet science. However, simply inferring planet formation processes from relative atmospheric abundances is a challenging task, due to the numerous complex processes which govern a planet's formation history \citep{vanderMarel_2021,Molliere_2022}. Although space-based observatories such as the \textit{James Webb Space Telescope} (JWST) and \textit{ARIEL} are expected to constrain these quantities to better than 0.2 dex \citep{Greene_2016}, they will be limited resources, and thus it is necessary to couple these results with high-resolution efforts, to gain a fuller understanding of these planetary systems. In fact, \cite{Line_2021} constrained the atmospheric C/O ratio of WASP-77Ab to within $\sim$~0.1 dex using high-resolution emission spectroscopy, emphasising the enormous potential of high-resolution retrievals, even in the era of JWST.  \newline \newline \indent In Section \ref{sec:2} we present the ESPRESSO observations and data reduction, the removal of spectral distortions in the form of the Centre-to-Limb (CLV) variation and Rossiter-McLaughlin (RM) effect, and the removal of stellar and telluric contamination with \textsc{SysRem}. In Section \ref{sec:3} we introduce our forward model atmosphere, and its subsequent model filtering, as well as demonstrating our analysis methods. These methods include the cross-correlation technique, as well as outlining our likelihood mapping approach, which is then folded into a full atmospheric retrieval framework. In Section \ref{sec:4} we present our results, as well as the search for potential exospheric species via the artificial ``scaling up'' of model absorption features. Finally, in Section \ref{sec:5}, we discuss our findings and present possible avenues to explore in future work, before concluding the study and summarising our results in Section \ref{sec:6}.

\section{ESPRESSO Observations and Data Reduction}
\label{sec:2}
We observed two full primary transits of WASP-121b with the ESPRESSO spectrograph \citep{Pepe_2021} at the VLT on the nights of 2021 January 26 and 2021 March 4, as part of the program 106.21R1 (PI: Gibson). ESPRESSO is the extremely stable, fibre-fed, high-resolution, cross-dispersed echelle spectrograph installed at the VLT. Installed at the inherent combined Coud\'{e} focus (ICCF), ESPRESSO is capable of illumination from any one of the unit telescopes (UT) at a time (1-UT mode), as well as all 4 unit telescopes simultaneously (4-UT mode), via the UT Coud\'{e} trains. Both transits were observed using the 1-UT mode, achieving a resolution of R$\sim$140,000. An archival partial primary transit of WASP-121b is also included in our analysis, observed in the 4-UT mode (R$\sim$70,000) on the night of 2018 November 30 during the commissioning of ESPRESSO, as part of the program 60.A-9128, previously analysed in \cite{Borsa_2020}. All three data sets cover a wavelength range of approximately 3770$-$7900 $\Angstrom$. The observed data sets are hereafter labelled T1, T2, and T3 in chronological order. Further details regarding our observations are given in Table~\ref{tab:Table1}.
\begin{table*}
	\centering
	\caption{A summary of the ESPRESSO observations of WASP-121b}
	\label{tab:Table1}
	\begin{tabularx}{\textwidth}{YYYYYYYY} 
		\hline
		\hline
		Data Set & Date & Observing Mode & Instrument Mode & N$_{\rm exp}$ (in-transit) & Exp. Time & S/N@550nm & Seeing \\
		\hline
		\hline
		T1 & 30-11-18 & 4-UT & MR42 & 29 (16) & 300 s & $\sim$150 & 0.7$''$ -- 1.1$''$\\
		T2 & 26-01-21 & 1-UT & HR11 & 96 (33) & 200 s & $\sim$19 & 0.79$''$ -- 1.35$''$\\
		T3 & 04-03-21 & 1-UT & HR11 & 93 (33) & 200 s & $\sim$15 & 0.79$''$ -- 2.15$''$\\
		\hline
	\end{tabularx}
\end{table*}
\newline \indent The data were reduced using the ESPRESSO Data Reduction Software (DRS) v2.3.3, and the two-dimensional non-blaze-corrected S2D spectra were extracted. The DRS includes standard reduction procedures for echelle time-series spectra, such as bias and dark subtraction, flat-field and contaminated pixel correction, as well as wavelength calibration and barycentric velocity correction.
\subsection{Spectral pre-processing}
\label{sec:2_1}
The extracted spectra were ``cleaned'' by subtracting a model from each spectral order. The model was constructed via the outer product of the spectral median (i.e. median over time) and temporal median (i.e. median over wavelength) of each order, weighted by the mean of the spectral median. A 10th order polynomial was then fitted to each spectrum of the residual orders, and any 5$\sigma$ outliers were replaced by their respective polynomial value, before adding the cleaned residual array back to the model. This procedure replaced approximately 0.05\%, 0.03\%, and 0.03\% of pixels for T1, T2, and T3, respectively.\newline \indent In order to optimise the information extracted from the time-series spectra, good initial estimates of the time- and wavelength-dependent uncertainties are required. We followed the procedure outlined in \citet{Gibson_2020}, summarised briefly here. An initial Poisson noise estimate of the form $\sigma_i = \sqrt{aF_i + b}$ was assumed for each pixel, where $F_i$ is the respective pixel flux value, and $a$ and $b$ are the noise scale factor and gain, respectively. A 5th order Principle Component Analysis (PCA) decomposition of the cleaned residual array was then constructed and subtracted, removing the stellar spectrum, resulting in a new residual array, $R$. Values for $a$ and $b$, and thus $\sigma_i$, were then found via the optimisation of a Gaussian log-likelihood of the form:
\begin{align*}
    \ln\mathcal{L}(a,b) = -0.5\sum_i\left({\frac{R_{i}}{\sigma_i}}\right)^2 - \sum_i\ln\sigma_i
\end{align*}
The uncertainties are constructed from the best-fit values for $a$ and $b$, before finally fitting this estimate with another 5th order PCA model, in order to remove biases in the Poisson noise estimate. This model is our final estimate of the uncertainties.\newline \indent The resultant spectral orders were then blaze-corrected following \citet{Merritt_2020}. Firstly, each order was divided by its mean spectrum. The resultant array was then smoothed with a median filter with a width of 11 pixels (to remove outliers) and a Gaussian filter with a standard deviation of 200 pixels. To ensure these values did not significantly distort the underlying exoplanet signal, we performed injection tests with an atmospheric model containing neutral Fe, Mg, and Cr, for each data set. The model had a negative $K_{\rm p}$, to ensure that the injected signal is separated from the real exoplanet signal. We then performed our retrieval analysis on the injected data, similar to that outlined in Section \ref{sec:Retrieval}, and found the retrieved model parameters, as well as the retrieved relative abundances and  $T$-$P$ profiles to be in agreement with the injected values. These results are outlined in Figs. \ref{fig:T1_Inj_tests}, \ref{fig:T2_Inj_tests}, and \ref{fig:T3_Inj_tests} for T1, T2, and T3, respectively. The original orders and their respective uncertainties were then divided by the resultant smoothed blaze function. This does not remove the blaze distortion, but instead places each order on a common blaze in time. This procedure also removes the low-frequency ``wiggle'' pattern seen in ESPRESSO spectra \citep{Allart_2020}, which has a period of $\sim$30 $\Angstrom$ and amplitude of $\sim$1$\%$ at 600 nm \citep{Tabernero_2021,Kesseli_2022}. There is also a second set of ``wiggles'' which affect ESPRESSO spectra, with a shorter period of $\sim$1 $\Angstrom$, and amplitude of $\sim$0.1$\%$ at 600 nm, which was minimised by using a narrower median filter width in our blaze correction, compared to our previous works. These wiggles are likely caused by an interference pattern induced by a combination of readout electronics and inhomogeneities in the Coud\'{e} train optics \citep{Sedaghati_2021}\footnote{The ESPRESSO consortium and ESO are working to characterise this effect \citep{Allart_2020}.}$^{,}$\footnote{See section 4.3 of the ESPRESSO user manual for details (\url{https://www.eso.org/sci/facilities/paranal/instruments/espresso/ESPRESSO_User_Manual_P109_v2.pdf}).}. After cleaning and blaze correction, the spectra were then Doppler shifted to the stellar rest frame via the stellar radial velocity, computed via the orbital parameters outlined in Table~\ref{tab:Table2}.
\begin{table}
	\centering
	\caption{Physical and orbital parameters of the WASP-121 system}
	\label{tab:Table2}
	\begin{tabular}{ccl} 
		\hline
		\hline
				Parameter & Value & Reference\\
		\hline
		\hline
    \multicolumn{1}{c}{} & \multicolumn{1}{c}{\hspace{1.0cm}\textit{Stellar parameters}}\\ \hline \\[-0.2cm]
    $R_\star$ & 1.458$\pm$0.030 R$_{\odot}$& \citet{Delrez_2016}\\[0.1cm]
    $M_\star$ & 1.353$^{+0.080}_{-0.079}$ M$_{\odot}$& \citet{Delrez_2016}\\[0.1cm]
    $T_{\rm eff}$ & 6460$\pm$140 K & \citet{Delrez_2016}\\[0.1cm]
    Spectral type & F6V & \citet{Delrez_2016}\\[0.1cm]
    $\log{g}$ [cgs] & 4.242$^{+0.011}_{-0.012}$ & \citet{Delrez_2016}\\[0.1cm]
    [Fe$/$H] & 0.13$\pm$0.09 & \citet{Delrez_2016}\\[0.1cm]
    $K_\star$ & 177.0$^{+8.5}_{-8.1}$ m s$^{-1}$ & \citet{Bourrier_2020}\\[0.1cm]
    $v_\star\sin{i}$ & 13.56$^{+0.69}_{-0.68}$ km s$^{-1}$ &
    \citet{Delrez_2016}\\[0.1cm]
    $v_{\rm sys}$ & 38.198$\pm$0.002 km s$^{-1}$ & \citet{Borsa_2020}\\[0.1cm]
    \hline

    \multicolumn{1}{c}{} &\multicolumn{1}{c}{\hspace{1.0cm}\textit{Planetary parameters}}\\ \hline \\[-0.2cm]
    $P$ & 1.27492504$^{+1.5\cdot10^{-7}}_{-1.4\cdot10^{-7}}$ days & \citet{Bourrier_2020}\\[0.1cm]
    $R_{\rm p}$ & 1.753$\pm$0.036 R$_{\rm Jup}$& \citet{Bourrier_2020}\\[0.1cm]
    $M_{\rm p}$ & 1.157$\pm$0.070 M$_{\rm Jup}$& \citet{Bourrier_2020}\\[0.1cm]
    $g_{\rm p}$ & 9.33$^{+0.71}_{-0.67}$ m s$^{-2}$ & \citet{Bourrier_2020}\\[0.1cm]
    $R_{\rm p}$/$R_\star$ & 0.12355$^{+0.00033}_{-0.00029}$ & \citet{Bourrier_2020}\\[0.1cm]
    $a_{\rm p}$/$R_\star$ & 3.8131$^{+0.0075}_{-0.0060}$& \citet{Bourrier_2020} \\[0.1cm]
    $i_{\rm p}$ & 88.49$\pm$0.16 $^{\circ}$ & \citet{Bourrier_2020} \\[0.1cm]
    $\lambda$ & 87.20$^{+0.41}_{-0.45}$ $^{\circ}$ & \citet{Bourrier_2020}\\[0.1cm]
    \hline
    \end{tabular}

\end{table}
\begin{figure}
	\includegraphics[width=\columnwidth]{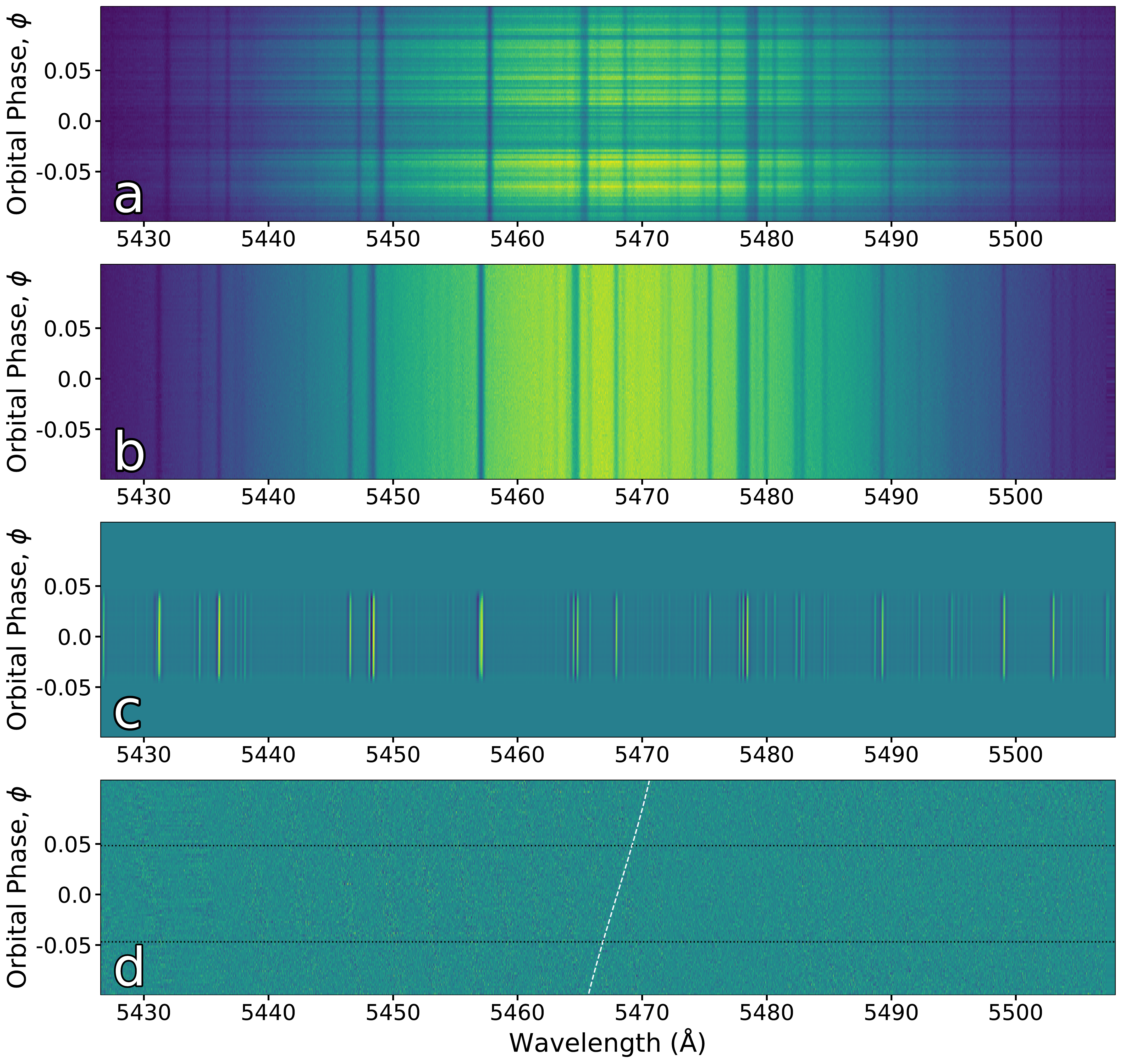}
    \caption{The data reduction steps applied to a single order of T2. \textbf{(a)} Cleaning and outlier removal. \textbf{(b)} Blaze correction and Doppler shift to the stellar rest frame. \textbf{(c)} The modelled CLV and RM distortions on the in-transit spectra. \textbf{(d)} After CLV and RM removal, and removal of stellar and telluric features with \textsc{SysRem}. The dotted horizontal lines show the orbital phases of ingress and egress, whereas the white dashed curve shows the approximate velocity shift of WASP-121b.}
    \label{fig:example_figure}
\end{figure}
\subsection{Removal of CLV and RM Effect}
In addition to contamination, or spectral distortion such as the blaze, imprinted on the spectra, there is also spectral imprints due to the spatial variation of the stellar spectrum across the stellar disk, whilst a planet is in transit, such as the Centre-to-Limb variation (CLV) and Rossiter-McLaughlin (RM) effect. As has become common practice in the reduction of high-resolution transmission spectroscopy observations \citep{Yan_2017,Casasayas_Barris_2017,Casasayas_Barris_2018,Yan_2019,Turner_2020,Nugroho_2020}, we modelled and removed these distortions of the stellar spectral lines, caused by both the CLV and RM effect, from the data. \newline \indent Following \citet{Nugroho_2020}, we generated synthethic stellar spectra at 21 different limb angles ($\mu = \cos{\theta}$) using \textsc{Spectroscopy Made Easy} (SME, \citealt{Piskunov_2017}), with linelists obtained from the VALD3\footnote{\url{http://vald.astro.uu.se}} database \citep{Ryabchikova_2015} using MARCS stellar models \citep{Gustafsson_2008}. A $\mu$ value was then calculated for each pixel on a stellar grid of radius 500 pixels, onto which the $\mu$-dependent synthetic spectra were linearly interpolated. A rotational velocity value was also calculated for each pixel, with the corresponding spectrum then Doppler shifted, assuming solid body rotation and ignoring gravity darkening. As the planet transits the stellar disk, the occulted pixels were masked using a binary mask, and the remaining spectra were integrated over the entire disk. This assumes the planet is an opaque disk of radius $R_{\rm p}$ across our entire wavelength range. For each exposure, a model stellar spectrum is generated, which includes the distortions from both the CLV and RM effect in-transit. We then divided the model by the out-of-transit spectrum, ensuring only the CLV and RM effect's distortions remain. Our data was then divided by this final model. Without this step, an anomalous ``Doppler shadow'' is apparent in the cross-correlation maps ($\Delta v_{\rm sys}\simeq 0$ km s$^{-1}$) of species present in the stellar atmosphere (see Fig. \ref{fig:RMCLV_CC_maps}). This anomaly was not prominent in our UVES observations of WASP-121b previously outlined in \cite{Gibson_2020,Gibson_2022} and \cite{Merritt_2020,Merritt_2021}, and due to its small change in velocity relative to WASP-121b in-transit, it was not modelled/removed from these observations (see \citealt{Gibson_2020} for further details).

\subsection{Removal of stellar and telluric features}
\label{sec:SysRem}
With the CLV and RM distortions of the stellar lines modelled and removed, the stellar and telluric lines are then (quasi-)stationary in time throughout the data. In order to remove the stationary spectral lines, \textsc{SysRem} \citep{Tamuz_2005} was applied to each order of the data. \textsc{SysRem} was first applied to high-resolution spectroscopy by \citet{Birkby_2013}, and has since become a common practice for the removal of stellar and telluric features in both transmission and emission spectra \citep{Nugroho_2017,Nugroho_2020,Nugroho_2020b,Birkby_2017,Gibson_2020,Gibson_2022,Merritt_2020,Serindag_2021}. Before applying \textsc{SysRem}, each order was divided by its median spectrum, in order to normalise the data. For each \textsc{SysRem} iteration, each spectral order is decomposed into two column vectors, $\textbf{u}$ and $\textbf{w}$, where the \textsc{SysRem} model for each order is given by their outer product, $\textbf{uw}^{\rm T}$. This model is then subtracted from the input array(s), to get the processed array(s), and the procedure is repeated on the processed array(s) for the next iteration.  Thus, the \textsc{SysRem} model for a single order, $S_{N}$, given after $N$ iterations is: 
\begin{align}
 S_{N} &= \sum_{i=1}^{N}  \textbf{u}_i \textbf{w}_i^{\rm T} = \textbf{U} \textbf{W}^{\rm T} 
 \label{eq:S_n}
\end{align}
where \textbf{U} and \textbf{W} are matrices containing column vectors $\textbf{u}_i$ and $\textbf{w}_i$. 
This final \textsc{SysRem} model is then subtracted from the normalised data, $R$:
\begin{align*}
 R_{N} &= R - S_{N}
\end{align*}
An alternative and arguably more accurate approach is to use \textsc{SysRem} to model the stellar/telluric spectrum directly, without first dividing each order by its median spectrum, and then divide the model through \citep{Gibson_2020} -- preserving the relative line depths, which is important for atmospheric retrievals. However, dividing through by the median spectrum in each order and subsequently subtracting the \textsc{SysRem} model allows for a faster model filtering process in each likelihood calculation, and we find little difference between these approaches.\newline \indent Cross-correlation and likelihood mapping analysis was performed for an atmospheric model with multiple species (Fe I, Mg I, Cr I, Ti I, V I, Na I, and Ca I), and the optimum number of \textsc{SysRem} iterations for each data set was determined from the resultant detection significance, conditioned on the optimum $K_{\rm p}$ and $\Delta v_{\rm sys}$ values, outlined in Section \ref{sec:4.1}. \citet{Nugroho_2017} separate each order and subsequently determine the optimum number of \textsc{SysRem} iterations for each order via injection tests, whereas we choose to keep the number of \textsc{SysRem} iterations constant across all orders. The approach of \citet{Nugroho_2017} benefits from the fact that the systematics vary across orders, which is then accounted for by order-based \textsc{SysRem} optimisation. Also, \textsc{SysRem} acts to distort the underlying exoplanet signal, the extent of which is reduced via this approach. However, for a fast model filtering framework (see Section \ref{sec:model_filtering}), it is more convenient to keep the number of \textsc{SysRem} iterations constant across all orders. Fixing the number of \textsc{SysRem} iterations also ensures that we don't bias the detection by optimising over each order, which may cause an overfitting of the stellar/telluric spectrum. The optimum number of \textsc{SysRem} iterations for each data set was 18, 1, and 4 for T1, T2, and T3, respectively.
\section{Methods}
\label{sec:3}
\subsection{Modelling the transmission spectra}
\label{sec:irradiator}
When performing atmospheric retrievals with high-resolution data sets, the computational speed of the underlying forward model, as well as how accurately it represents the exoplanet's signal, is critical. For our analysis we used a 1D transmission spectrum model, \textsc{irradiator}, described in detail in \citet{Gibson_2022}, computed across a wavelength range of 3670$-$8000 $\Angstrom$, at a constant resolution of R~=~200,000. This model utilises pressure-dependent absorption cross-sections, allowing us to recover the atmosphere's vertical temperature structure whilst also providing more reliable abundance constraints. Here we assume a well-mixed atmosphere. Within our atmospheric retrieval framework, hundreds of thousands of forward models are generated at high-resolution across a broad spectral range, thus the aim of this forward model is primarily computational speed, whilst also preserving the accuracy of the underlying transmission spectrum.\newline \indent In our analysis, we include absorption cross-sections for neutral atomic species (Fe I, Mg I, Cr I, Ti I, V I, Na I, and Ca I), all of which were previously detected in the atmosphere of WASP-121b at high-resolution, with the exception of Ti I \citep{Gibson_2020, Hoeijmakers_2020, Borsa_2020, Ben_Yami_2020}, using the precomputed opacity grids provided by petitRADTRANS\footnote{\url{https://petitradtrans.readthedocs.io/en/latest/}} \citep{Molliere_2019}. We define 70 atmospheric layers, equidistant in log space, ranging from $10^2$ bar to $10^{-12}$ bar, across which we compute a $T$-$P$ profile using the parametric model from \citet{Guillot_2010} with input parameters $T_{\rm irr}$, $T_{\rm int}$, $\kappa_{\rm IR}$, and $\gamma$ (see Eq. 29 of \citet{Guillot_2010} for reference), corresponding to the irradiation temperature, internal temperature, mean infrared opacity, and the ratio of visible-to-infrared opacity, respectively. After computing the vertical extent of the atmosphere assuming hydrostatic equilibrium, and accounting for varying temperature and gravity in each layer, we solve the radiative transfer equations by computing the opacity of each predefined layer of our atmosphere, weighing each species' cross-sections on their respective volume mixing ratios (VMRs), $\chi_{\rm species}$, and integrating through the grazing geometry. The computed optical depth for each layer is then converted to an effective planetary radius. We also mask any values below the corresponding radius of a given cloud deck pressure, $P_{\rm cloud}$, to account for an opaque cloud deck. In addition, we include a H$_2$ VMR, $\chi_{\rm ray}$, along with H$_2$ opacities from \citet{Dalgarno_1962}, to account for Rayleigh scattering, with the minimum prior limit corresponding to Jupiter's H$_2$ abundance. The truncated transmission spectrum is then converted from an effective planetary radius to a differential flux ($\Delta F$) value.
\subsection{Model filtering}
\label{sec:model_filtering}
As mentioned above, how well our model transmission spectra represent the underlying exoplanet signal is critical when performing model comparison within Bayesian inference algorithms, in order to retrieve accurate constraints of atmospheric parameters. As outlined in Section \ref{sec:SysRem}, data preprocessing techniques -- such as \textsc{SysRem} -- distort the exoplanet signal, and thus it is important to apply these same processing steps to our forward model, such that it more accurately represents the data. However, \textsc{SysRem} is an iterative process, and therefore the execution time is too long to be practically folded into a retrieval framework. Thus, \citet{Gibson_2022} introduced a novel model filtering technique, which utilises the output matrices \textbf{U} and \textbf{W}, containing the column and row vectors \textbf{u} and \textbf{w} for each \textsc{SysRem} iteration (see Section \ref{sec:SysRem}). Firstly, our forward model computed above is broadened via convolution with a Gaussian kernel, and linearly interpolated to the 2D wavelength grid of our data (order $\times$ wavelength). For each order, the 2D forward model is then Doppler shifted to a planetary velocity given by:
\begin{align}
 \textbf{v}_{\rm \textbf{p}} &= K_{\rm p}\sin{(2\pi\cdot \bm{\phi} )} + \Delta v_{\rm sys}
 \label{eq:planet_vel}
\end{align}
where $K_{\rm p}$, $\bm{\phi}$, and $\Delta v_{\rm sys}$ are the planet's radial velocity semi-amplitude, orbital phase, and the systemic velocity offset, respectively.
This results in a 3D shifted forward model (time/phase $\times$ order $\times$ wavelength).\newline \indent In Eq. \ref{eq:S_n}, $\textbf{U} \textbf{W}^{\rm T}$ can be seen as a linear basis model, where \textbf{U} contains $N$ basis vectors (for each iteration of \textsc{SysRem}), and \textbf{W} the corresponding weights. In practice, we fit \textbf{U} to each order, \textbf{M} (time/phase $\times$ wavelength), of the forward model, using a linear least squares to find the best-fit weights:
\begin{align}
 \textbf{w}^\mathbf{\prime} &= (\textbf{U}^{\rm T}\textbf{U})^{-1}\textbf{U}^{\rm T} \textbf{y}
\end{align}
for a single column vector \textbf{y}, or:
\begin{align}
 \textbf{W}^\mathbf{\prime} &= (\textbf{U}^{\rm T}\textbf{U})^{-1}\textbf{U}^{\rm T} \textbf{Y}
 \label{eq:W'}
\end{align}
for a 2D time $\times$ wavelength array \textbf{Y}. The best-fit model, \textbf{Y}$^\mathbf{\prime}$, to the array \textbf{Y}, is then given by the outer product of the best-fit weights, \textbf{W}$^\mathbf{\prime}$, and the basis vectors \textbf{U}, where the term $(\textbf{U}^{\rm T}\textbf{U})^{-1}\textbf{U}^{\rm T}$, in Eq. \ref{eq:W'}, is the Moore-Penrose inverse, $\textbf{U}^{\dagger}$, such that:
\begin{align}
 \textbf{Y}^\mathbf{\prime} &= \textbf{U}\textbf{U}^{\dagger} \mathbf{Y}
\end{align}
However, to account for the data uncertainties, which were accounted for initially when computing \textbf{U}, we take the mean of the uncertainties over wavelength, $\boldsymbol{\skew{-.5}\hat\sigma}$, for each order, as only the relative weights will affect the best-fit, and it is too slow to provide each array of uncertainties for every wavelength. Therefore, the final fit to the forward model is:
\begin{align}
 \textbf{M}^\mathbf{\prime} &= \textbf{U}(\boldsymbol{\Lambda}\textbf{U})^{\dagger}\left(\boldsymbol{\Lambda} \textbf{M}\right)
 \label{eq:filter}
\end{align}
where $\boldsymbol{\Lambda}$ is a diagonal matrix of 1$/\boldsymbol{\skew{-.5}\hat\sigma}$ terms. \textbf{M}$^\mathbf{\prime}$ is then subtracted from the forward model, \textbf{M}, equivalent to the treatment of the data with the \textsc{SysRem} model in Section \ref{sec:SysRem}. The term $\textbf{U}(\boldsymbol{\Lambda}\textbf{U})^{\dagger}\boldsymbol{\Lambda}$ depends solely on the \textsc{SysRem} basis vectors and the data uncertainties, meaning it can be precomputed, allowing Eq. \ref{eq:filter} to be applied directly for each given forward model. This procedure is computed independently for each order, each with their own basis vectors, in each likelihood calculation of our retrieval, and allows for the same preproccessing steps to be applied to the model, as the data, without the expensive computational overhead of iterative processes such as \textsc{SysRem}. This is also considerably faster than PCA, where \textbf{U} is also recomputed for each iteration.  
\subsection{Cross-correlation}
\label{sec:CC}
\begin{figure}
	\includegraphics[width=\columnwidth]{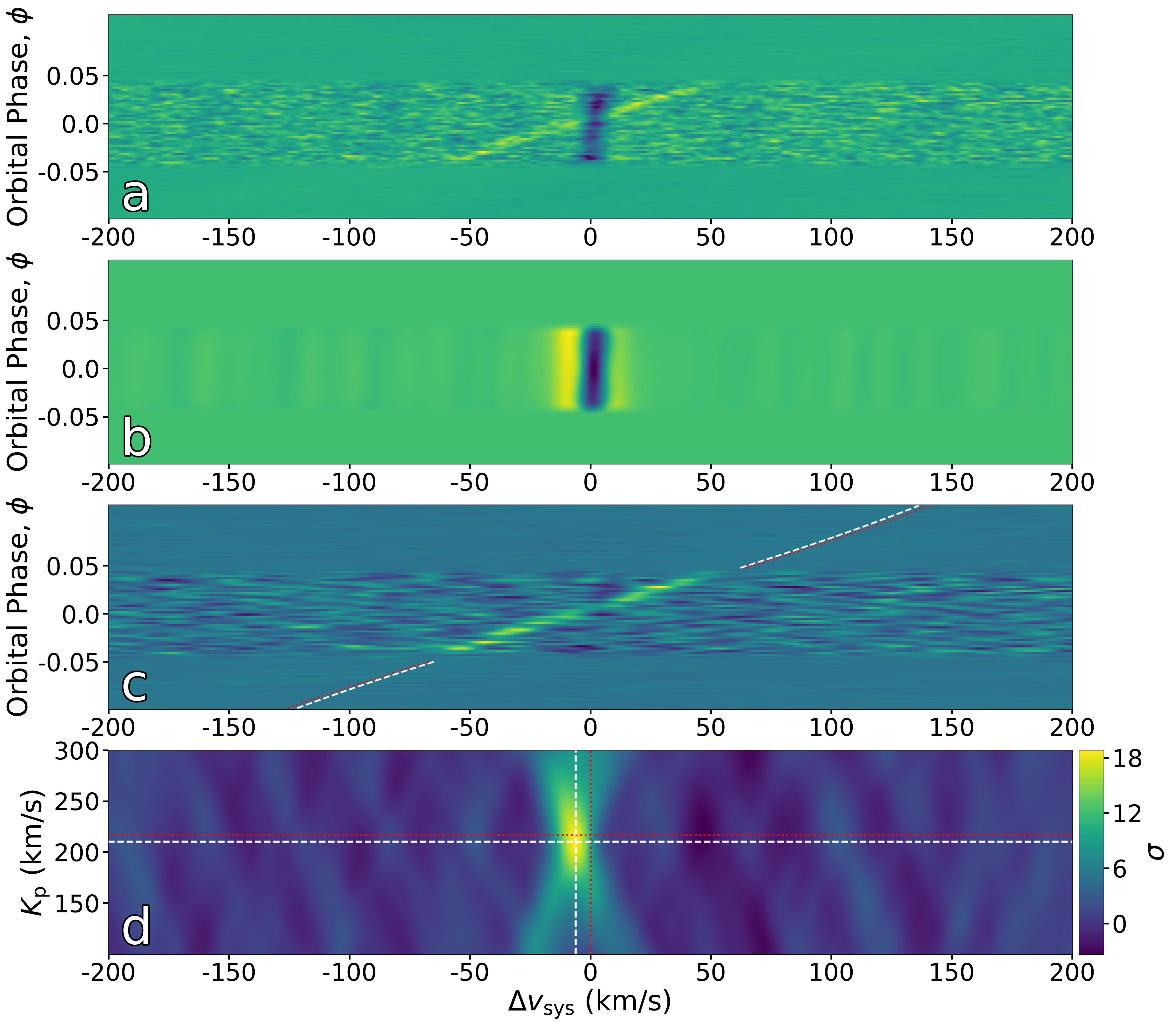}
    \caption{The planetary signal detected via cross-correlation analysis of T2 with our combined (Fe I, Mg I, Cr I, Ti I, V I, Na I, and Ca I) atmospheric model. \textbf{(a)} Cross-correlation map before the removal of CLM and RM distortions, weighted proportional to the transit lightcurve. \textbf{(b)} The ``Doppler shadow'' produced by cross-correlating the CLM and RM distortions with our model. \textbf{(c)} Cross-correlation map after the removal of CLM and RM distortions. The red dotted curve shows the expected planetary velocity trail from RV measurements, whereas the white dashed curve shows the planetary velocity trail corresponding to the maximum $K_{\rm p}$ and $\Delta v_{\rm sys}$ values. \textbf{(d)} The final $K_{\rm p}$--$\Delta v_{\rm sys}$ map, showing an 18.8$\sigma$ detection near the expected planetary velocity.}
    \label{fig:RMCLV_CC_maps}
\end{figure}

Equipped with our data, free of spectral distortions and stellar and telluric contamination, as well as our Doppler shifted model, filtered to mimic the effect of our data preprocessing steps, we are ready to perform traditional cross-correlation analysis (e.g. \citealt{Snellen_2010,Ehrenreich_2020,Gibson_2020,Hoeijmakers_2020,Merritt_2020,Merritt_2021}).  The data and the shifted model are multiplied, before summing their products over both wavelength and spectral order, in order to produce a cross-correlation function (CCF). This is mathematically equivalent to a dot product, such that:
\begin{align}
\rm{CCF} (\Delta v_{\rm sys}) &= \sum_{i=1}^{N} \frac{f_i m_i (\Delta v_{\rm sys})}{\sigma_i^2}
 \label{eq:CC}
\end{align}
where the product is weighted on the variance of the data, $\sigma_i^2$, and the summation occurs over both wavelength and spectral order. {Eq.~\ref{eq:CC}} produces cross-correlation values for each combination of orbital phase, $\phi$, and $\Delta v_{\rm sys}$, via Eq. \ref{eq:planet_vel}, referred to as a “$\phi$--$\Delta v_{\rm sys}$” map, or simply, a “cross-correlation” map. \newline \indent The cross-correlation map in Fig. \ref{fig:RMCLV_CC_maps} shows a visible trail of the planetary signal. However, for weaker signals in which this trail is not visible, and also to constrain the planetary orbital velocity, it is necessary to integrate the cross-correlation maps over a range of predicted planetary velocities, $v_{\rm p}$. In practice, a range of radial velocity semi-amplitude values, $K_{\rm p}$, are chosen, in the vicinity of the predicted value from RV measurements. The cross-correlation function for each orbital phase is then shifted to a new planetary velocity given by Eq. \ref{eq:planet_vel} for a given $K_{\rm p}$, and integrated as a function of time/phase, “collapsing” the signal in time. The collapsed array is stored for each value of $K_{\rm p}$, producing a velocity-summed cross-correlation map, known as a “$K_{\rm p}$--$\Delta v_{\rm sys}$” map. In order to place constraints on the detection significance of a given species within a $K_{\rm p}$--$\Delta v_{\rm sys}$ map, we subtract the map by the mean of the noise in regions away from the peak, before dividing the map by the standard deviation of the noise away from the peak \citep{Brogi_2012,Brogi_2018}. However, the regions away from the peak in which the mean/standard deviation of the noise is calculated are arbitrarily chosen, and thus the resultant detection significance is not exact. Therefore, one of the limitations of this technique is the conversion of the cross-correlation values into a statistically robust goodness-of-fit estimate, such as a chi-squared statistic or likelihood value, for direct model comparison \citep{Brogi_2016,Brogi_2017}.
\subsection{Likelihood mapping}
\label{sec:likelihood_mapping}
\citet{Brogi_Line_2019} first introduced a method to ``map'' cross-correlation values of a given atmospheric model to a likelihood value. An alternative, yet similar method was developed by \citet{Gibson_2020} using a full Gaussian likelihood function, which allows for time- and wavelength-dependent uncertainties to be accounted for. We will briefly outline this method here, for a detailed description see \citet{Gibson_2020}. Beginning with a full Gaussian likelihood function, with uncertainties that vary in time and wavelength:
\begin{align}
  \mathcal{L}(\boldsymbol{\theta}) = \prod_{i=1}^{N} \frac{1}{\sqrt{2\pi(\beta\sigma_i)^2}} \cdot\exp\bigg(-\frac{1}{2}\frac{(f_i - \alpha m_i(\boldsymbol{\theta}))^2}{\beta\sigma_i^2}\bigg)
 \label{eq:L1}
\end{align}
where $\alpha$ and $\beta$ are a model scale factor and a noise scale factor, respectively. $\boldsymbol{\theta}$ represents a vector of model parameters. As opposed to Eq.~\ref{eq:CC}, $i$ now refers to wavelength, spectral order, \textit{and} time. Dropping reference to $\boldsymbol{\theta}$, the natural logarithm of the likelihood, or log-likelihood, is then computed as follows:
\begin{align}
    \ln{\mathcal{L}} = -\frac{N}{2}\ln{2\pi} - \sum_{i=1}^{N}\ln{\sigma_i} - N\ln{\beta} - \frac{1}{2}\chi^2
        \label{eq:7}
\end{align}
where
\begin{align}
        \chi^2 = \sum_{i=1}^{N}\frac{(f_i - \alpha m_i)^2}{(\beta\sigma_i)^2}
        \label{eq:chi2}
\end{align}
The first two terms in Eq. \ref{eq:7} are constant for a given data set, and thus can be dropped, giving:
\begin{align}
    \ln{\mathcal{L}} = - N\ln{\beta} - \frac{1}{2}\chi^2
    \label{eq:finL}
\end{align}
Expanding Eq. \ref{eq:chi2} gives:
\begin{align}
    \label{eq:10}
    \chi^2 = \frac{1}{\beta^2}\Bigg(\sum_{i=1}^{N}\frac{f_i^2}{\sigma_i^2} + \alpha^2\sum_{i=1}^{N}\frac{m_i^2}{\sigma_i^2} -2\alpha\sum_{i=1}^{N}\frac{f_im_i}{\sigma_i^2}\Bigg)
\end{align}
The final summation in Eq. \ref{eq:10} is equivalent to the CCF (Eq. \ref{eq:CC}), summed over time, outlined in Section \ref{sec:CC}, such that:
\begin{align}
        \chi^2 = \frac{1}{\beta^2}\Bigg(\sum_{i=1}^{N}\frac{f_i^2}{\sigma_i^2} + \alpha^2\sum_{i=1}^{N}\frac{m_i^2}{\sigma_i^2} -2\alpha  \hspace{0.1cm}\textrm{CCF}\Bigg)
        \label{eq:chi22}
\end{align}
Eqs. \ref{eq:finL} and \ref{eq:chi22} allow for the computation of the log-likelihood directly from the CCF, and thus the ability to perform model comparison and direct atmospheric retrievals from high-resolution observations whilst utilising the power of the cross-correlation method.

\subsection{Atmospheric Retrieval}
\label{sec:Retrieval}
In order to perform a full retrieval analysis on our data sets, we must compute a forward model for a given set of model parameters $\boldsymbol{\theta}$. $\boldsymbol{\theta}$ is comprised of $\{\alpha$, $\beta$, $K_{\rm p}$, $\Delta v_{\rm sys}$, W$_{\rm conv}\}$, where W$_{\rm conv}$ is the width of the Gaussian broadening kernel in pixels, as well as the \textsc{irradiator} input parameters $\{T_{\rm irr}$, $T_{\rm int}$, $\kappa_{\rm IR}$, $\gamma$, $P_{\rm cloud}$, $\chi_{\rm ray}$, $\chi_{\rm species}$\} (see Section~\ref{sec:irradiator}). \newline \indent Adopting uniform prior distributions, each spectrum of the forward model is divided by the its median value, to imitate the blaze correction, and then filtered, as described in Section~\ref{sec:model_filtering}. The log-posterior is computed by computing the log-prior and adding the log-likelihood (Eq. \ref{eq:finL}) for a given set of model parameters. This can then be folded into an MCMC framework, in order to sample the posterior, and retrieve an estimate of the posterior distributions of the model parameters for each data set. We use a Differential-Evolution Markov Chain (DEMC) \citep{TerBraak_2006,Eastman_2013}, running an MCMC chain with 192 walkers,  with a burn-in length of 200 and a chain length of 300, resulting in 96,000 samples of the posterior, of which 38,400 are discarded. In each of our retrievals, $T_{\rm int}$ is fixed at 200 K, and we fit for the remaining parameters. Fixing $T_{\rm int}$ at 200 K should have a minimal influence on the $T$-$P$ structure of UHJs observed via transmission spectroscopy, where we are sensitive to lower pressures. The chains' convergence is checked via the Gelman-Rubin statistic, after splitting each chain into four separate subchains. We also divide the chains into groups of independent walkers, and overplot the 1D and 2D marginal distributions in different colours, in order to visualise the convergence of our MCMC (see Fig. \ref{fig:T2_retrieval}).

\section{Analysis}
\label{sec:4}
\subsection{Cross-correlation and Likelihood mapping}
\label{sec:4.1}
Cross-correlation analysis, outlined in Section \ref{sec:CC}, was performed with a model transmission spectrum, containing neutral Fe, Mg, Cr, Ti, V, Na, and Ca, for each of our reduced observations, T1, T2, and T3, using the best-fit model parameters from our atmospheric retrieval. A cross-correlation map before the removal of the CLV and RM distortions in-transit is shown in the upper panel of Fig. \ref{fig:RMCLV_CC_maps}. The anomalous ``Doppler shadow'' produced by these spectral distortions after cross-correlation with a model template is shown in the second panel of Fig. \ref{fig:RMCLV_CC_maps}, with the final ``corrected'' cross-correlation map of our model template shown in the third panel of Fig. \ref{fig:RMCLV_CC_maps}. This corrected cross-correlation map was then shifted to a range of planetary velocities and collapsed over time, producing a $K_{\rm p}$--$\Delta v_{\rm sys}$ map, shown in the bottom panel of Fig. \ref{fig:RMCLV_CC_maps}. We compute a formal detection significance of 40.5$\sigma$, 18.8$\sigma$, and 11.4$\sigma$ for our best-fit atmospheric models,  as outlined in Section \ref{sec:CC}, for the T1, T2, and T3 data sets, respectively. It is worth noting that the significance of these values is difficult to interpret statistically, as they are sensitive to arbitrary choices. We also compute the detection significance of each best-fit model using the likelihood distribution of our model scale factor, $\alpha$, conditioned on the optimum $K_{\rm p}$ and $\Delta v_{\rm sys}$ values, similar to the approach of \cite{Gibson_2020}. We extract the mean and standard deviation of $\alpha$ from this conditional distribution, and divide the mean by the standard deviation to get a detection significance. This method, therefore, does not rely on different regions of the $K_{\rm p}$--$\Delta v_{\rm sys}$ map when calculating the noise properties and thus the detection significance, as before. Using this method, we calculate a detection significance of 71.3$\sigma$, 23.9$\sigma$, and 15.2$\sigma$ for the T1, T2, and T3 data sets, respectively. While it is difficult to interpret and compare these values, it is clear that we get strong detections of our atmospheric models for each of the data sets.
\begin{figure*}
    \includegraphics[width=\textwidth]{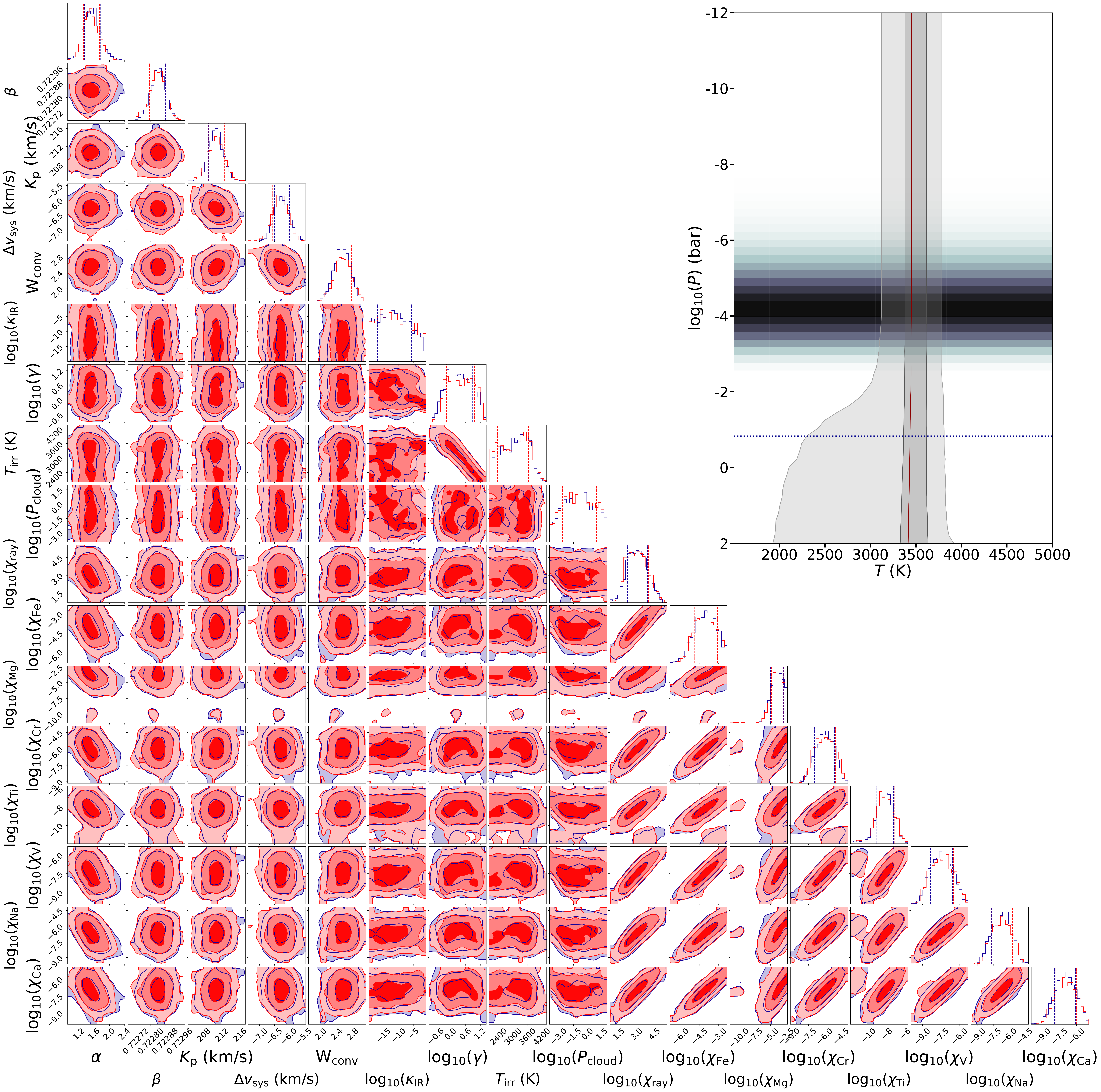}
    \caption{A summary of our retrieval results for T2, with the 1D and 2D marginalised posterior distributions of each of our model parameters displayed within a corner diagram. The red and blue posterior distributions represent independent subchains, of the same MCMC chain, both converging to similar distributions. The correlations in the retrieved abundances are evident in the lower right panels of the corner diagram. The $T$-$P$ profile on the right was computed from 10,000 random samples of the MCMC, where the red curve shows the median profile, and the grey and lighter grey shaded regions show the 1$\sigma$ and 2$\sigma$ contours, respectively. The blue horizontal dotted line is the median retrieved cloud deck pressure, $\log_{10}(P_{\rm cloud})$, below which the model spectrum is truncated. Also plotted is the mean contribution function for the ESPRESSO wavelength range. This corner diagram was generated using \textsc{corner} \citep{corner}.}
    \label{fig:T2_retrieval}
\end{figure*}
\subsection{Relative abundance constraints}
\begin{figure*}
	\includegraphics[width=\textwidth]{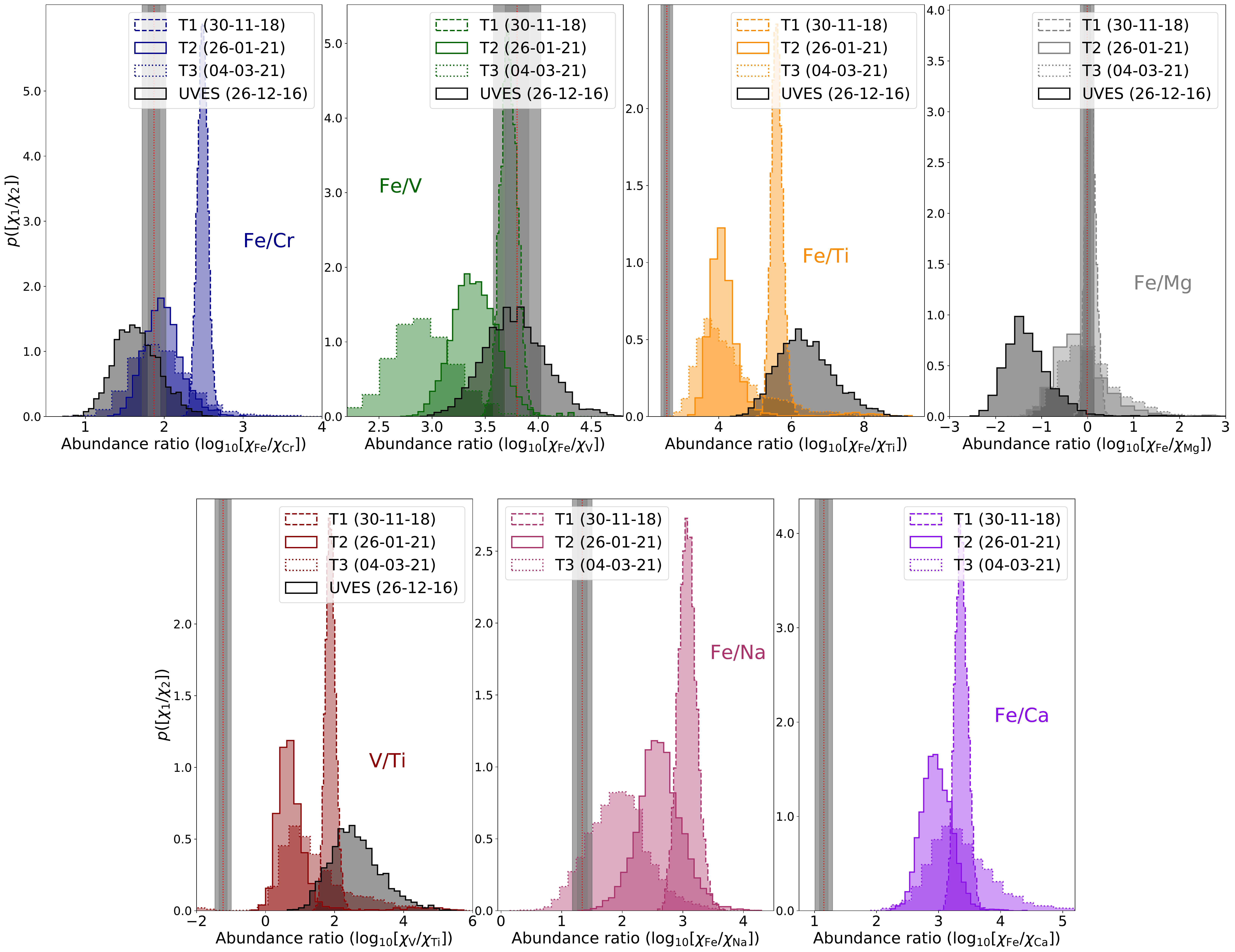}
    \caption{Comparison of the relative abundance constraints from our atmospheric retrievals for each of our observations. The vertical red dotted lines, and the grey shaded regions, represent the relative abundance ratio of WASP-121, and their respective 1$\sigma$ and 2$\sigma$ contours, calculated using elemental abundances from \protect\cite{Polanski_2022} and solar elemental abundances from \protect\cite{Asplund_2009}. The literature solar values are $\log_{10}(\chi_{\rm Fe}/\chi_{\rm Cr}) = 1.86\pm0.06$,  $\log_{10}(\chi_{\rm Fe}/\chi_{\rm V}) = 3.57\pm0.09$,  $\log_{10}(\chi_{\rm Fe}/\chi_{\rm Ti}) = 2.55\pm0.06$,  $\log_{10}(\chi_{\rm Fe}/\chi_{\rm Mg}) = -0.10\pm0.06$, $\log_{10}(\chi_{\rm V}/\chi_{\rm Ti}) = -1.02\pm0.09$, 
    $\log_{10}(\chi_{\rm Fe}/\chi_{\rm Na}) = 1.26\pm0.06$, and
    $\log_{10}(\chi_{\rm Fe}/\chi_{\rm Ca}) = 1.16\pm0.06$
    \protect\citep{Asplund_2009}.}
    \label{fig:rel_abuns}
\end{figure*}

The retrieved parameters from our ESPRESSO observations T1, T2, and T3 were then compared, alongside a previous transit of WASP-121b observed on the night of 2016 December 26 with the UVES spectrograph (see \citealt{Gibson_2020,Gibson_2022,Merritt_2020,Merritt_2021} for further details regarding this data set). A summary of the retrieval results for T2 are shown in Fig. \ref{fig:T2_retrieval}, whereas the retrieval results for T1, and T3 are given in the Appendix (Figs. \ref{fig:T1_retrieval} and \ref{fig:T3_retrieval}, respectively). \newline \indent Our retrieval framework also allows us to constrain the absolute abundances of each species, however, our transmission spectra lack an absolute normalisation, resulting in broad constraints on the absolute abundances, with significant correlations. The relative abundances, however, are constrained more precisely (see Fig. \ref{fig:rel_abuns}) for each of our observations, from the relative line strengths of each species.\newline \indent The retrieved relative abundance constraints for each data set is shown in Fig. \ref{fig:rel_abuns}. The relative abundance constraints retrieved from our UVES observations \citep{Gibson_2022} are shown as black histograms, however neutral Na and Ca were not included in this analysis, whereas we have included them here.

\subsection{The search for exospheric species}
\begin{figure*}
    
	\includegraphics[width=\textwidth]{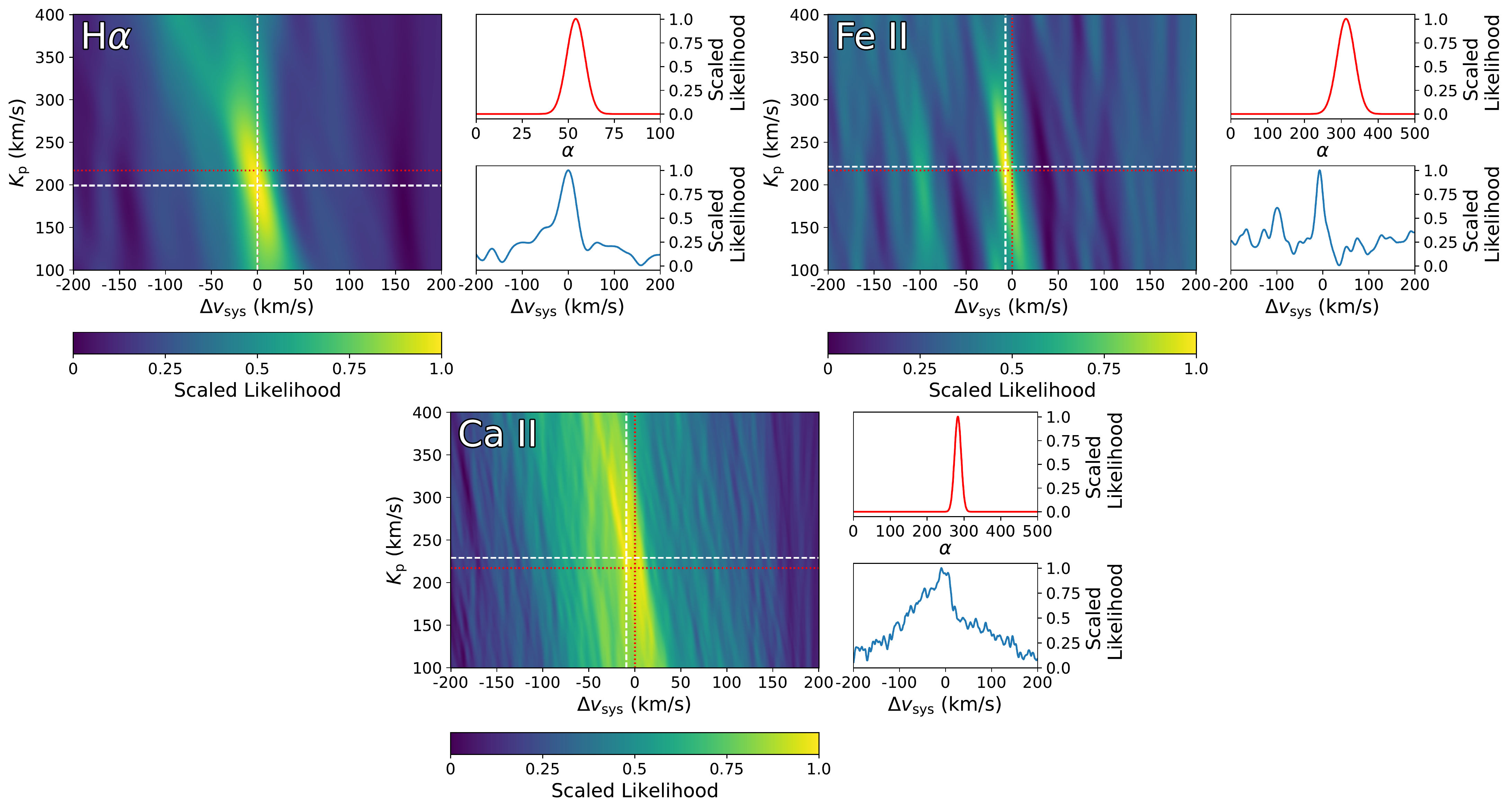}
    \caption{The likelihood maps and conditional likelihood distributions of $\alpha$ for the potential exospheric species, for T1 only. (\textit{left subpanel}) The maximum likelihood map, normalised between 0 and 1 for ease of visualisation. The red dotted lines shows the expected planetary velocity values, whereas the white dashed lines shows the maximum $K_{\rm p}$ and $\Delta v_{\rm sys}$ values. (\textit{upper right subpanel}) The conditional distribution of $\alpha$ (red), normalised between 0 and 1, fitted with a Gaussian function (blue). (\textit{lower right subpanel}) The likelihood function at the maximum $K_{\rm p}$ value, outlining the various $\Delta v_{\rm sys}$ values for each species. The multi-peaked, ``diamond-like'' shape of the Ca II signal may also be evidence of Ca II present in two different atmospheric limbs of the planet}
    \label{fig:Exo_lnL}
\end{figure*}

\begin{figure}
    
	\includegraphics[width=\columnwidth]{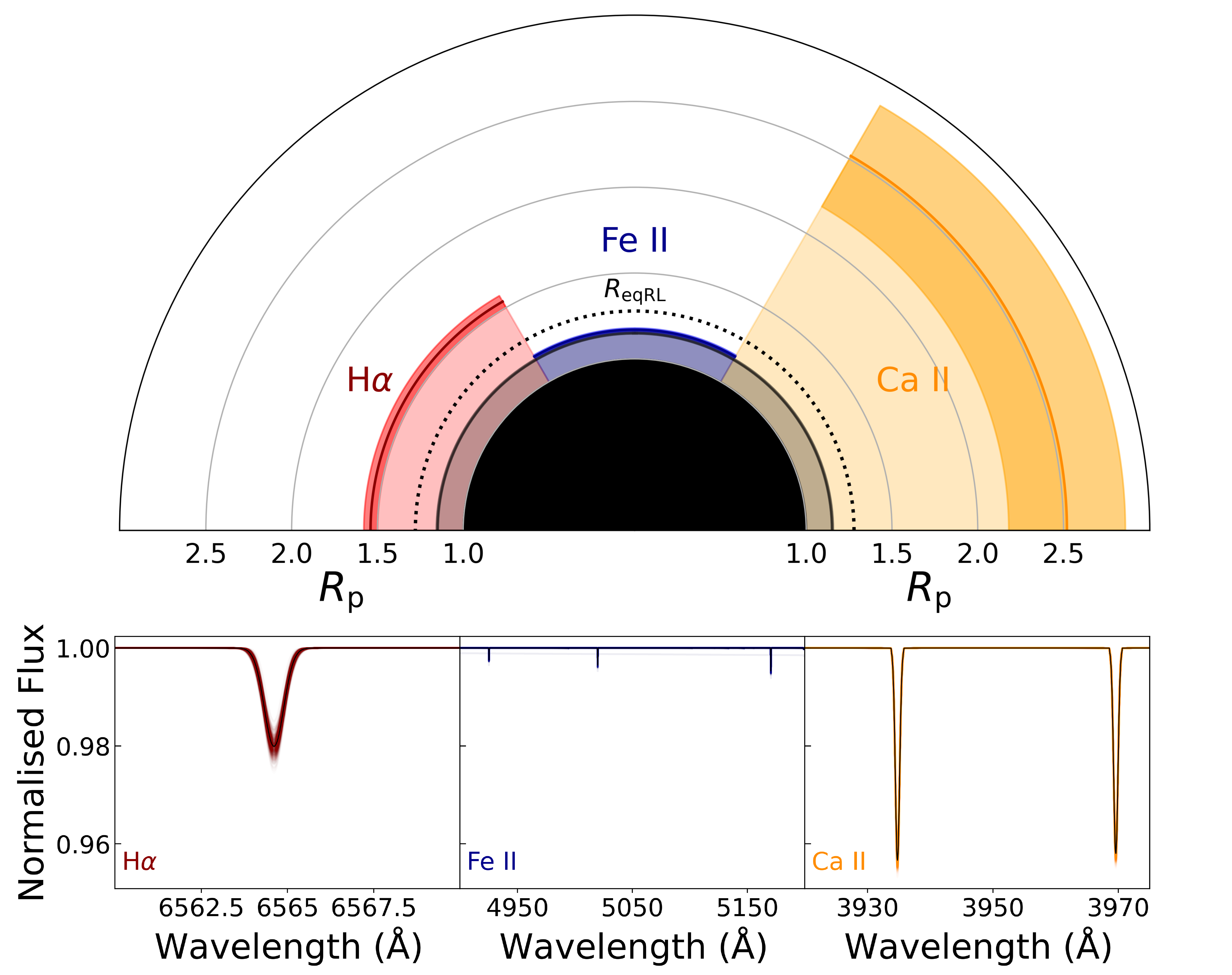}
    \caption{(\textit{upper}) A schematic diagram showing the extent of potential exospheric species via the artificial ``scaling up'' of their transmission spectra with our model scale factor, $\alpha$. The dark shaded regions represent the $1\sigma$ contours. The transit equivalent Roche limit, $R_{\rm eqRL}$, is shown as a black dotted line. The black shaded region highlights the maximum vertical extent of the~best-fit combined atmospheric model for T1 (see Fig. \ref{fig:T1_retrieval}). This diagram is not indicative of any real latitudinal variation in the extent of these species. (\textit{lower}) Scaled model transmission spectra, generated from 10,000 random samples of the MCMC, centered around the strongest line(s) for each of the exospheric species. The median transmission spectra are shown in black}
    \label{fig:Exo_Rp_comp}
\end{figure}

 H$\alpha$, Fe II, Mg II, and Ca II have previously been reported to exist at high altitudes in the exosphere of WASP-121b. These recent detections motivated our search for potentially exospheric species, using the likelihood mapping method outlined in Section \ref{sec:likelihood_mapping}. This search was conducted solely using the T1 data set, due to the higher S/N and lower resolution data, resulting in a faster likelihood evaluation. Although our underlying atmospheric model assumes hydrostatic equilibrium, an assumption which ultimately breaks down in a hydrodynamically outflowing envelope, we can map the existence of these species via the artifical ``scaling up'' of our model absorption features via our model scale factor, $\alpha$. As outlined in Fig. \ref{fig:T2_retrieval}, retrieving an $\alpha$ value close to 1 reaffirms the assumption of hydrostatic equilibrium, which is valid for species lower in the atmosphere. However, retrieving a relatively large $\alpha$ value, orders of magnitude greater than 1, would be explained by species in an extended upper atmosphere. For this reason, these species were excluded from previous retrievals, due to a varying $\alpha$ value between species in different layers of the atmosphere.
\newline \indent
We performed a full atmospheric retrieval for each individual exospheric species, similar to that outlined in Section \ref{sec:Retrieval} -- with the exception of Mg II, which we were unable to detect in this data set. The best-fit model parameters were used to generate model transmission spectra of H$\alpha$, Fe II, and Ca II individually, from which we compute a likelihood map and a conditional likelihood distribution of $\alpha$, shown in Fig. \ref{fig:Exo_lnL}. We then drew 10,000 random samples of the MCMC, for each individual species, and generated model transmission spectra. From this, we computed the median and standard deviation of the scaled transmission spectra, and found the resultant effective planetary radius values, $R_{\rm p}$, from the \textit{strongest absorption lines} of each transmission spectra. From this we can map the extent of these absorption signals in the upper atmosphere (see Fig. \ref{fig:Exo_Rp_comp}). The scaled transmission spectra result in absorption features extending up to $1.54\pm0.04$ $R_{\rm p}$ , $1.17\pm0.01$ $R_{\rm p}$, and $2.52\pm0.34$ $R_{\rm p}$, for H$\alpha$, Fe II, and Ca II, respectively.  

\section{Discussion}
\label{sec:5}
 The cross-correlation analysis outlined above clearly detects our combined transmission spectrum model containing neutral Fe, Mg, Cr, V, Ti, Na, and Ca, with a detection significance of 40.5$\sigma$, 18.8$\sigma$, and 11.4$\sigma$ for the T1, T2, and T3 data sets, respectively. Our transmission spectrum model was produced using \textsc{irradiator} \citep{Gibson_2022}, outlined in Section \ref{sec:irradiator}, which accounts for pressure-dependent cross-sections, Rayleigh scattering in the form of H$_2$ opacities combined with a H$_2$ VMR, and an opaque cloud deck. This result was produced after the removal of CLV and RM distortions from our ESPRESSO data sets. Our forward model was then filtered using a novel model filtering technique \citep{Gibson_2022}, which acts to reproduce the distortion of iterative algorithms such as \textsc{SysRem} on the underlying observed exoplanet signal. The cross-correlation value can also be mapped to a likelihood value, outlined in Section \ref{sec:likelihood_mapping}, and a direct likelihood evaluation then enables a full retrieval framework. Thus, we may constrain the absolute abundances of each individual species, as well as the velocity shifts, $T$-$P$ structure, line broadening parameters, and vertical extent of our atmosphere.\\ \newline \indent From our $T$-$P$ profile constraints (Fig. \ref{fig:T2_retrieval}), the temperature in the upper atmosphere, at pressures $>$ 10$^{-2}$ bar, is approximately 3450$\pm$160 K, which is consistent within 2$\sigma$ across all data sets, as well as with previous UVES observations of WASP-121b \citep{Gibson_2020,Gibson_2022}. We retrieve significantly higher temperatures, however, than those inferred for the nightside of the planet from low-resolution studies \citep{Bourrier_2020b,Evans_2018,Wilson_2021,Mikal_Evans_2022}, and the dayside temperature of 2870$\pm$50 K inferred from TESS phase curve measurements \citep{Bourrier_2020b}. We also retrieve similar temperatures to the dayside temperature of $\sim 3200$ K, obtained by \cite{Mikal_Evans_2022}. This could be explained by high-resolution studies probing higher up in the atmosphere, where we would expect higher temperatures due to the previously reported thermal inversion \citep{Evans_2017}, which is hinted at by the existence of ionised species in the upper atmosphere. The existence of ionised species at high altitudes, however, could also be explained by a lower electron pressure, which acts to significantly ionise, as well as photoionisation at lower pressures. Deeper in the atmosphere, we do not constrain the $T$-$P$ structure well, which is expected, where emission studies would be more sensitive to the temperature structure of the deeper atmosphere. It is important to note that transmission studies are typically not very sensitive to the $T$-$P$ structure of the atmosphere, despite the stringent constraints, and therefore we must be careful not to overinterpret our retrieved values.\\ \newline \indent The constrained atmospheric dynamic parameters, such as $K_{\rm p}$ and $\Delta v_{\rm sys}$, from which we calculate the planetary orbital velocity, $v_{\rm p}$, are also consistent across all observations, with a net blueshift of $\Delta v_{\rm sys}{\sim}$ 6 km s$^{-1}$ for our combined model, similar to previous results from high-resolution transmission spectroscopy studies of WASP-121b \citep{Gibson_2020,Gibson_2022,Hoeijmakers_2020,Merritt_2020,Merritt_2021,Borsa_2020}. The constraints on the planetary orbital velocity, $v_{\rm p}$, are outlined in Fig. \ref{fig:vp_comp}, where it is evident that this quantity is well constrained in-transit. We also constrain the width of a Gaussian broadening kernel, W$_{\rm conv}$, to 2.82$\pm$0.07 pixels, 2.54$\pm$0.20 pixels, and 1.50$\pm$0.68 pixels, for T1, T2, and T3, respectively. For an atmospheric model computed at a resolution of R~=~200,000, or velocity of $\approx$~1.5 km s$^{-1}$, these broadening kernel widths convert to full widths at half maximum (FWHM) of 9.89$\pm$0.25 km s$^{-1}$, 8.91$\pm$0.70 km s$^{-1}$, and 5.26$\pm$2.39 km s$^{-1}$, for T1, T2, and T3, respectively. As the equatorial rotational velocity of WASP-121b is 7.4 km s$^{-1}$ ($\approx$~15 km s$^{-1}$ peak-to-peak), assuming it's tidally locked, our measurements of the broadening do not indicate clear evidence of broadening beyond rotation in the lower atmosphere. However, this analysis would benefit from a full modelling of planetary wind patterns.  \\ \newline \indent 
Due to the degeneracies between the chosen reference transit radius and the reference pressure of the atmosphere, our transmission spectra lack an absolute normalisation \citep{Heng_2017}, limiting our ability to place stringent constraints on the \textit{absolute} abundance of each of our model species. The constraints on the absolute abundances are quite broad, with strong correlations, which can be seen in the lower right panels of the corner diagram in Fig. \ref{fig:T2_retrieval}. However, the \textit{relative} abundances can be constrained much more precisely, as although we lose information regarding the absolute abundances, the relative abundance is constrained from the relative line strengths, assuming a well-mixed, hydrostatic atmosphere \citep{Benneke_2012}. \\ \newline \indent The retrieved relative abundance ratios were compared with the relative abundance ratios of WASP-121, calculated using elemental abundances from \cite{Polanski_2022} and solar elemental abundances from \cite{Asplund_2009}. In the upper left panel of Fig. \ref{fig:rel_abuns}, the retrieved abundance ratio $\log_{10}(\chi_{\rm Fe}/\chi_{\rm Cr})$ is consistent across each of our ESPRESSO observations within 1$\sigma$, and is also consistent with stellar values for our T2, T3, and UVES observations, whereas T1 deviates from stellar. For $\log_{10}(\chi_{\rm Fe}/\chi_{\rm V})$, we retrieve consistent values across all observations, with T1 and T3 consistent within 2$\sigma$. These values also agree with stellar values, within 1$\sigma$, however T3 is consistent within 2$\sigma$ . Fe and V have been previously detected in transmission in the atmosphere of WASP-121b \citep{Gibson_2020,Hoeijmakers_2020}, and have a relatively large number of spectral lines in our observed wavelength range, which makes their relative abundances an ideal target for this analysis. The $\log_{10}(\chi_{\rm Fe}/\chi_{\rm Mg})$ relative abundance ratio is also in agreement across all observations, within 1$\sigma$, and is consistent with stellar values. The relative abundance ratios which include Ti, $\log_{10}(\chi_{\rm Fe}/\chi_{\rm Ti})$ and $\log_{10}(\chi_{\rm V}/\chi_{\rm Ti})$, are broadly consistent between our observations, however are inconsistent with stellar, which can be explained by a non-detection of Ti in each of the data sets. Also, \cite{Evans_2018} proposed that TiO undergoes nightside condensation and rain-out, in order to explain the absence of TiO in their transmission spectrum. This would result in the depletion of Ti, as chemical equilibrium would drive the remaining Ti to TiO until all remaining Ti bearing species are condensed out \citep{Hoeijmakers_2020}. Thus, if Ti were present, we would expect the relative abundances, $\log_{10}(\chi_{\rm Fe}/\chi_{\rm Ti})$ and $\log_{10}(\chi_{\rm V}/\chi_{\rm Ti})$, to be superstellar, as all Ti bearing species in the atmosphere of WASP-121b would have condensed out of the gas phase. The $\log_{10}(\chi_{\rm Fe}/\chi_{\rm Na})$ abundance ratio is in agreement across each of our ESPRESSO observations, within 1$\sigma$, with T3 consistent with stellar values. $\log_{10}(\chi_{\rm Fe}/\chi_{\rm Ca})$ is also consistent across each of our ESPRESSO observations, however is larger than stellar values. This could be explained by our model assuming a hydrostatic, well-mixed atmosphere, whereas these species are likely impacted by an escaping upper atmosphere, as Fe II and Ca II have previously been detected at high altitudes \citep{Sing_2019,Borsa_2020}.  $\log_{10}(\chi_{\rm Fe}/\chi_{\rm Ca})$ is significantly larger than the reported stellar value, however the absolute abundance of Fe is similar to the stellar value, whereas the absolute abundance of Ca is less than the reported stellar value. With Ca II extending to much greater altitudes than Fe II (Fig. \ref{fig:Exo_Rp_comp}), this result suggests that Ca could possibly undergo greater ionisation than Fe in the atmosphere of WASP-121b. A similar result could be inferred from the inconsistency of $\log_{10}(\chi_{\rm Fe}/\chi_{\rm Na})$ with stellar values for T1, T2, and UVES, where the absolute abundance of Na is also less than stellar, although Na II was not detected in this data set.\\ \newline \indent \cite{Mikal_Evans_2022} recently obtained an overall metallicity value of [M/H] $= 0.76\substack{+0.30 \\ -0.62}$, which is approximately 1$-$10$\times$ solar. This result, however, is difficult to directly compare to the results of our analysis, as in their study, the overall heavy metal enrichment is allowed to vary, whilst they fix the relative abundance ratios of the included species to solar values. It is important to note that our analysis would greatly benefit from a more sophisticated atmospheric model, in order to constrain and interpret our relative abundance values more precisely, and to directly compare with works such as \cite{Mikal_Evans_2022} and future results. In the future, we will look to perform similar analyses coupled with a model which assumes changing atmospheric chemistry with pressure/altitude, as well as a more complete treatment of atmospheric dynamics.   \\ \newline \indent We also probed the vertical extent of species previously reported at high altitudes in the atmosphere of WASP-121b, namely H$\alpha$, Fe II, and Ca II, using the T1 data set. \cite{Sing_2019} first detected exospheric Fe II and Mg II extending beyond the Roche lobe of WASP-121b in the UV using HST/STIS, at $\sim$2.7$R_{\rm p}$ and $\sim$2.5$R_{\rm p}$, respectively. This detection led to multiple follow-up campaigns at high-resolution, in search of potentially exospheric metal ions. Fe II was then detected in the optical with HARPS \citep{Ben_Yami_2020}. \cite{Cabot_2020} also detected H$\alpha$ with HARPS, up to $\sim$1.5$R_{\rm p}$, again attributed to an extended upper atmosphere. Finally, \cite{Borsa_2020} confirmed the detection of Fe II with ESPRESSO in the 1-UT mode, and also found both H$\alpha$ and Ca II, present beyond the Roche lobe of WASP-121b, at $\sim$ 1.4 $R_{\rm p}$ and $\sim$ 2 $R_{\rm p}$, respectively. Our results are summarised in Fig. \ref{fig:Exo_lnL}. Each exospheric species, with the exception of Mg II, was detected near the expected planetary velocity of WASP-121b, with each model spectrum requiring a significantly larger model scale factor, $\alpha$, than 1, which hints at their existence in an extended/outflowing upper atmosphere. This is similar to results obtained by \cite{Ben_Yami_2020}, where Fe II required a higher weighted absorption strength relative to species in the lower atmosphere (Fe I, Cr I, and V I). Similarly, \cite{Hoeijmakers_2020} detected neutral Mg in the atmosphere of WASP-121b, which required a significantly greater line amplitude in comparison to other species, which is consistent with our expectation of an outflowing atmosphere. We converted our ``scaled up'' transmission spectra to an effective planetary radius, displayed in Fig. \ref{fig:Exo_Rp_comp}, with H$\alpha$ extending to $1.54\pm0.04$ $R_{\rm p}$, Fe II to $1.17\pm0.01$ $R_{\rm p}$, and Ca II to $2.52\pm0.34$ $R_{\rm p}$, where the  transit equivalent Roche limit  $R_{\rm eqRL}$ $\sim$ 1.3 $R_{\rm p}$ \citep{Sing_2019}. Furthermore, there is a clear variability in the systemic velocity offset, $\Delta v_{\rm sys}$, and the width of the Gaussian broadening kernel, W$_{\rm conv}$, between our exospheric species. H$\alpha$, Fe II, and Ca II were detected at a $\Delta v_{\rm sys}$ of -1.8 km s$^{-1}$, -7.7 km s$^{-1}$, and -16.2 km s$^{-1}$, respectively. At a model resolution of R = 200,000, or velocity of $\approx$~1.5 km s$^{-1}$, the retrieved W$_{\rm conv}$ values, converted to a FWHM, are 26.95$\pm$2.01 km s$^{-1}$, 10.74$\pm$0.92 km s$^{-1}$, and 61.21$\pm$3.89 km s$^{-1}$, for H$\alpha$, Fe II, and Ca II, respectively. The variability in the velocity constraints of the exospheric species, relative to one another, as well as to species detected lower in the atmosphere ($\alpha{\simeq}$1), hints at species existing in different dynamical regions of the atmosphere \citep{Prinoth_2022}. The broad, ``diamond-like'' shape of the Ca II signal (see Fig. \ref{fig:Exo_lnL}) explains the significantly large broadening of Ca II, and  may be evidence of Ca II present in two different atmospheric limbs of the planet \citep{Nugroho_2020,Wardenier_2021}.\\ \newline \indent A long-standing goal of exoplanet studies is to link the observed bulk composition of an exoplanet to that planet’s formation pathway and evolution history. Previous studies have placed constraints on the atmospheric C/O ratio, which can be used to map the formation mechanism and migration pathway of HJ/UHJs \citep{Oberg_2011,Madhusudhan_2014}. Recent work by \cite{Lothringer_2021} has shown that refractory elements, such as Fe, Mg, and Si, can also be used to map the formation and migration pathways of UHJs, via their “refractory-to-volatile” ratio. These species are thought to condense into solids at T $<$ 2000 K, and thus are unobservable via transmission spectroscopy. However, in hotter objects with T $>$ 2000 K, such as WASP-121b, these species avoid condensation, such that constraints can be placed on their abundance, allowing their abundance ratios relative to volatile species, such as H$_2$O, CO, and CO$_2$, to be computed. The relative abundance constraints obtained in this study are solely for refractory species, with similar condensation temperatures, meaning that alone they are not ideal in tracing the formation history of the planet. However, the consistency and precision of these constraints, across timescales of months to years between transits, shows great promise for future measurements of relative abundances from high-resolution transmission spectroscopy, which can then be used to constrain the refractory-to-volatile ratio and infer planet formation and evolution processes. For example, $\log_{10}(\chi_{\rm Fe}/\chi_{\rm V})$ is constrained to 3.72$\pm$0.08, 3.40$\pm$0.21, and 2.98$\pm$0.51 for T1, T2, and T3, respectively. This is equivalent to 0.15$\pm$0.12 dex, -0.17$\pm$0.23 dex, and -0.59$\pm$0.52 dex, respectively, emphasising the importance of these
results alongside low-resolution efforts in the era of JWST.\\ \newline \indent The importance of the ``hierarchy of models'' \citep{Fortney_2021} used in retrievals of exoplanet transmission/emission spectra cannot be understated. With the ever increasing data quality provided by advanced space- and ground-based observatories, it is necessary to consider the three-dimensional nature of HJ/UHJs when modelling their spectra \citep{MacDonald_2022}. It has been shown that abundance gradients across the morning and evening limbs can result in biased inferences when using 1D models \citep{Pluriel_2022}, therefore caution must be taken when interpreting retrieved atmospheric properties from 1D retrievals, as these assumptions inherently result in non-negligible biases in the retrieved parameters. The computational cost of complex 3D forward models, particularly when folded into a retrieval framework with high-resolution data, must also be considered, in order to optimise the information content of the rich data sets on the horizon.\\ \newline \indent
 Recent discoveries outlining the spatial asymmetry in the atmospheres of HJ/UHJs must also be taken into account when constraining the atmospheric composition of these objects, with a longitudinal variation in the abundance of Fe I proposed in the atmosphere of WASP-76b \citep{Ehrenreich_2020, Kesseli_2021}. However, \cite{Savel_2022} have modelled the atmosphere of WASP-76b with three-dimensional Global Circulation Models (GCMs), in which a combination of Fe condensation, optically thick clouds, and a small nonzero eccentricity is needed to replicate the findings of \cite{Ehrenreich_2020}. \cite{Borsa_2020} also found variations in the velocity shift of the Fe I signal in the atmosphere of WASP-121b, between the beginning ($-2.80 \pm 0.28$ km s$^{-1}$) and end ($-7.66 \pm 0.16$ km s$^{-1}$) of transit, alluding to different dynamical regions being probed during transit. \cite{Wardenier_2021b} also showed that transmission spectroscopy studies of some UHJs, including WASP-121b, observe different atmospheric regions before and after mid-transit, due to the relatively large rotation angle of these planets during transit. \cite{Prinoth_2022} also found evidence for chemical inhomogeneity in the atmosphere of WASP-189b, using high-resolution cross-correlation spectroscopy, again emphasising the importance of 3D effects in interpreting the nature of UHJs. \\ \newline \indent In addition to the multidimensional nature of WASP-121b, the complex dynamics of an outflowing upper atmosphere must also be considered, and a full exospheric treatment with vertically outflowing winds could also be implemented, outlined in previous works using transmission spectroscopy at high-resolution \citep{Seidel_2020, Dos_Santos_2022}, in order to map the extent of these species, and infer the physical and dynamical properties of these regions.
\section{Conclusions}
\label{sec:6}
In this study, we presented multiple high-resolution transmission spectroscopy observations of the UHJ WASP-121b, using the ESPRESSO echelle spectrograph installed at the VLT, across several months/years. We demonstrated the removal of spectral distortions caused by the CLV and RM effect, as well as the removal of \mbox{(quasi-)}stationary stellar and telluric spectral lines with \textsc{SysRem}.
\begin{itemize}
    \item From these observations, we have retrieved consistent relative abundance constraints of neutral metals in the atmosphere of WASP-121b, which are broadly consistent with literature stellar values, across several high-resolution instruments/instrument modes, including previous relative abundance constraints obtained with VLT/UVES.\\
    \item We have also consistently constrained the $T$-$P$ structure of WASP-121b across multiple observations, which are in agreement with previously reported values at high-resolution.\\
    \item The atmospheric dynamic parameters, such as the planetary orbital velocity and broadening of spectral features are also constrained, and are consistent with previous results across each of our observations. \\
    \item  We have also implemented our retrieval framework to map the vertical extent of ionised metals which have been previously detected in the exosphere of WASP-121b, via our model scale factor, $\alpha$. This artificial scaling up resulted in the absorption features of these species extending to the similar high altitudes as reported in the literature. These exospheric species also show clear variability in their velocity constraints, which hints further at their existence in different dynamical regions of the atmosphere.
\end{itemize}
 
\indent This study is a strong validation of our model filtering and retrieval framework, as well as the stability of the atmosphere of WASP-121b over the observed timescales. Studies such as this, coupled with low-resolution efforts in the era of JWST and high-resolution efforts in the NIR/IR, as well as advanced atmospheric modelling, will place further constraints on these quantities, bridging the gap between the observed composition of these objects and their formation mechanism and evolution history.

\section*{Acknowledgements}
This work relied on observations collected at the European Organisation for Astronomical Research in the Southern Hemisphere. The author(s) gratefully acknowledge support from Science Foundation Ireland and the Royal Society in the form of a Research Fellows Enhancement Award. We are grateful to the developers of the NumPy, SciPy, Matplotlib, iPython, corner, petitRADTRANS, and Astropy packages, which were used extensively in this work \citep{Numpy,Scipy,matplotlib,ipython,corner,Molliere_2019,astropy}.

\section*{Data Availability}
The observations detailed in this publication are publicly available in the ESO Science Archive Facility (\url{http://archive.eso.org}) under program name 60.A-9128(H), 106.21R1.003, 106.21R1.002, and 098.C-0547. Data products will be shared on reasonable request.
 



\bibliographystyle{mnras}
\bibliography{example} 



\appendix
\section*{Appendix}
We have included additional plots regarding our injection tests detailed in Section \ref{sec:2_1} below. Additional 1D and 2D posterior distributions, similar to Fig. \ref{fig:T2_retrieval}, for T1 and T3 have also been included. A table of best-fit atmospheric model parameters found for each data set is also given in Table \ref{tab:TableA1} below.
\onecolumn


\renewcommand{\thefigure}{A\arabic{figure}}

\setcounter{figure}{0}
\begin{center}
    \begin{figure}
        \includegraphics[width=\columnwidth]{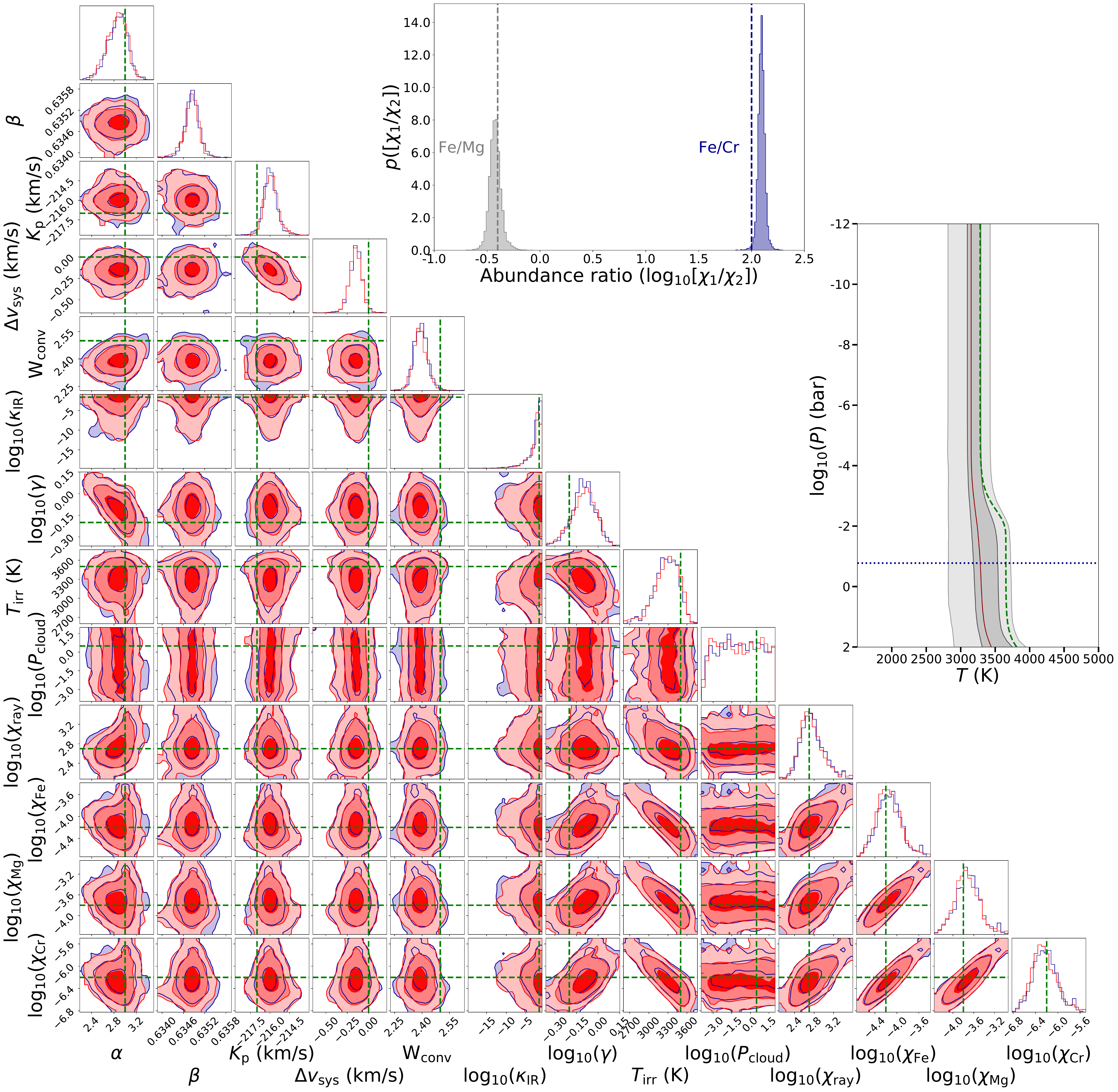}
        \caption{The results of our injection tests described in Section \ref{sec:2_1}, for T1. (\textit{left subpanel}) 1D and 2D marginalised posterior distributions of each of our model parameters, with the injected values shown as green dashed lines in each. (\textit{upper mid subpanel}) The retrieved relative abundance constraints, $\log_{10}(\chi_{\rm Fe}/\chi_{\rm Mg})$ and $\log_{10}(\chi_{\rm Fe}/\chi_{\rm Cr})$, and their injected values highlighted as vertical dashed lines. (\textit{right subpanel}) The retrieved $T$-$P$ profile, and its injected value shown as a green dashed curve. The blue horizontal dotted line is the median retrieved cloud deck pressure, $\log_{10}(P_{\rm cloud})$, below which the model spectrum is truncated.}
        \label{fig:T1_Inj_tests}
    \end{figure}
    \begin{figure}
        \includegraphics[width=\columnwidth]{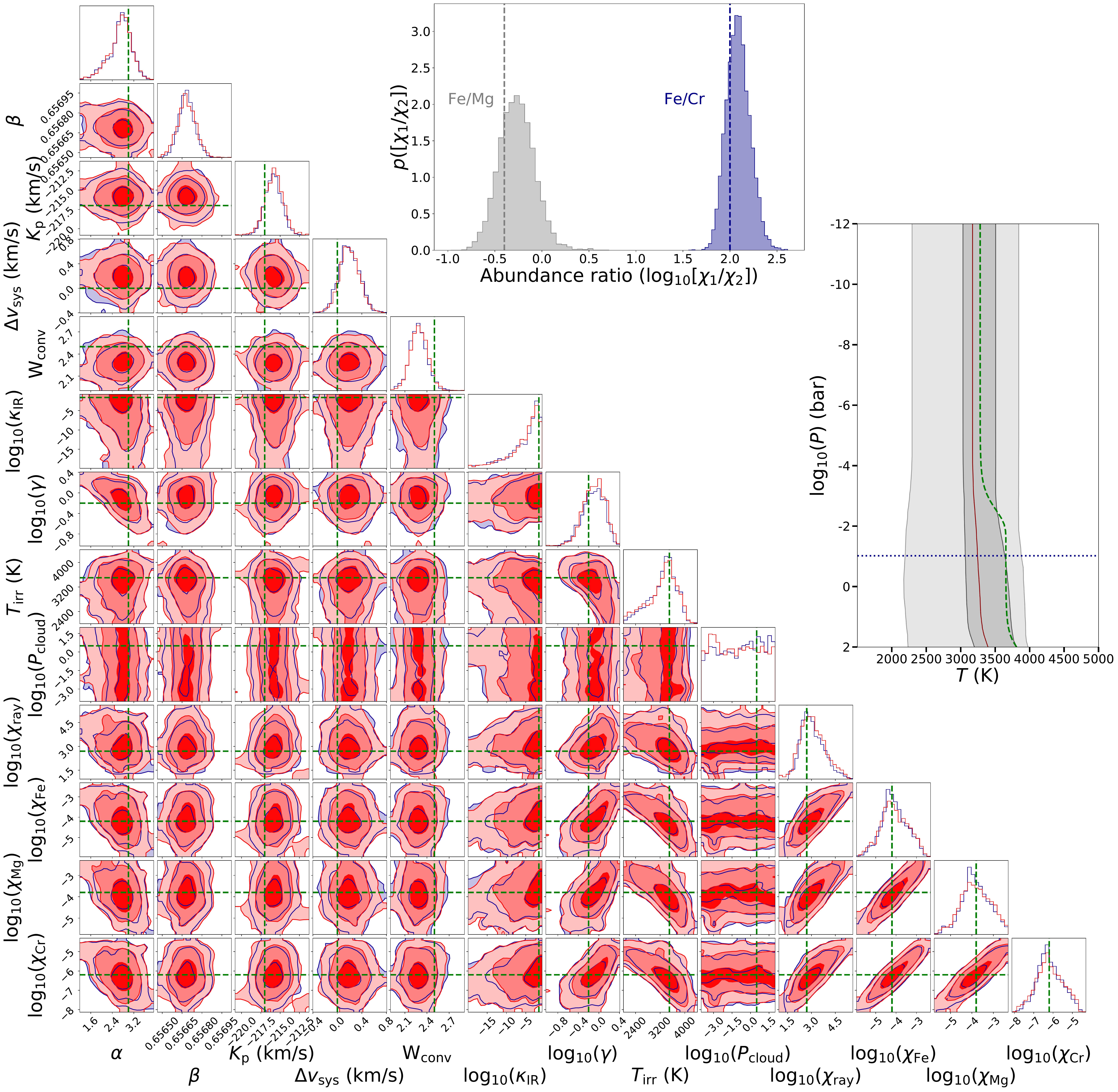}
        \caption{Similar to Fig. \ref{fig:T1_Inj_tests}, but for T2.}
        \label{fig:T2_Inj_tests}
    \end{figure}
    \begin{figure}
        \includegraphics[width=\columnwidth]{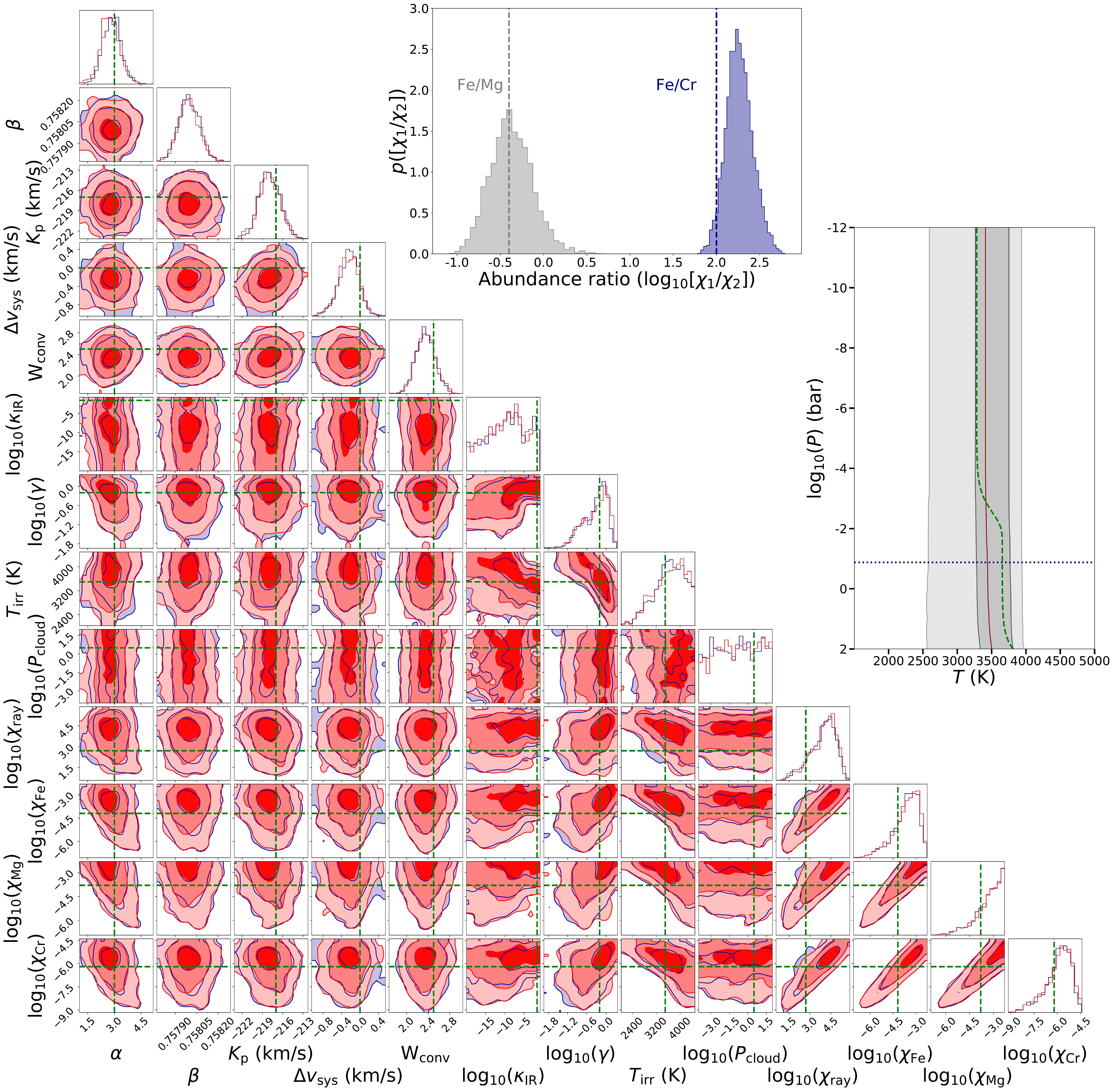}
        \caption{Similar to Fig. \ref{fig:T1_Inj_tests}, but for T3.}
        \label{fig:T3_Inj_tests}
    \end{figure}
    \begin{figure}
        \includegraphics[width=\columnwidth]{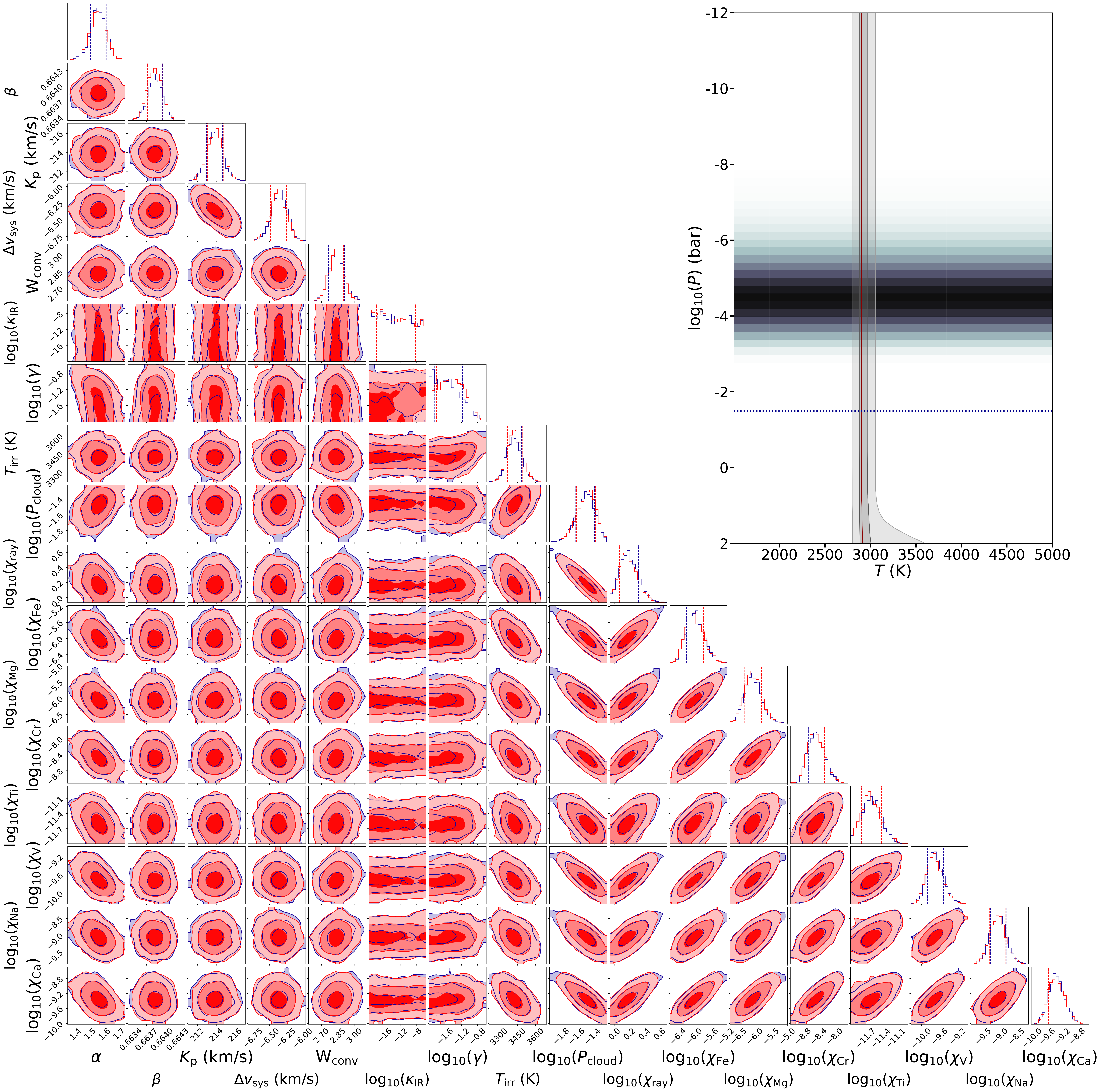}
        \caption{Similar to Fig. \ref{fig:T2_retrieval}, but for T1.}
        \label{fig:T1_retrieval}
    \end{figure}

    \begin{figure}
        \includegraphics[width=\columnwidth]{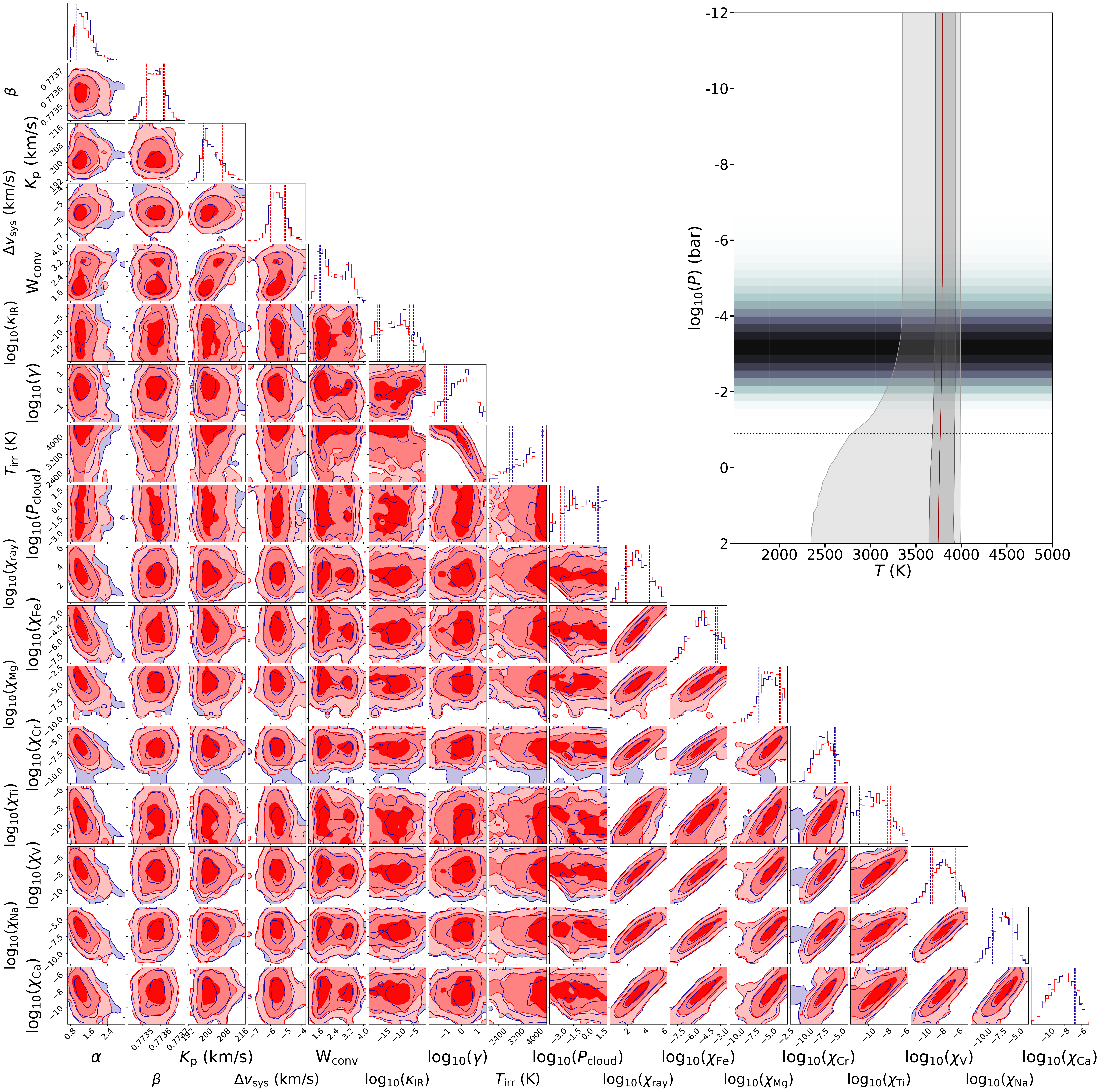}
        \caption{Similar to Fig. \ref{fig:T2_retrieval}, but for T3.}
        \label{fig:T3_retrieval}
    \end{figure}
\renewcommand{\thetable}{A\arabic{table}}

\setcounter{table}{0}
    \begin{table}
	\centering
	\caption{Best-fit model parameters for each of our atmospheric retrievals.}
	\label{tab:TableA1}
	\begin{tabular}{cccccc} 
		\hline
		\hline
				Parameter & & & T1 & T2 & T3\\
		\hline
		\hline
    $\alpha$ & & & 1.59$\pm$0.06  & 1.51$\pm$0.22 & 1.54$\pm$0.32\\[0.1cm]
    $\beta$ & & & 0.66$\pm$1.42$\cdot 10^{-4}$ & 0.72$\pm$4.23$\cdot 10^{-5}$& 0.77$\pm$4.72$\cdot 10^{-5}$\\[0.1cm]
    $K_{\rm p}$ (km s$^{-1}$)& & & 213.61$\pm$0.88 & 208.80$\pm$1.65 & 197.67$\pm$4.24\\[0.1cm]
    $\Delta v_{\rm sys}$ (km s$^{-1}$) & & & -6.35$\pm$0.12 & -6.26$\pm$0.26 & -5.92$\pm$0.45\\[0.1cm]
    W$_{\rm conv}$ (pixels) & & & 2.82$\pm$0.07 & 2.54$\pm$0.20 & 1.50$\pm$0.68 \\[0.1cm]
    $\log_{10}(\kappa_{\rm IR})$ & & & -9.48$\pm$4.18 & -1.58$\pm$5.18 & -15.92$\pm$4.80\\[0.1cm]
    $\log_{10}(\gamma)$ & & & -1.42$\pm$0.31 & -0.21$\pm$0.51 & -0.14$\pm$0.76\\[0.1cm]
    $T_{\rm irr}$ (K)& & & 3417.12$\pm$61.19 & 3688.82$\pm$573.53 & 3937.438$\pm$609.25 \\[0.1cm]
    $\log_{10}(P_{\rm cloud})$ (bar)& & & -1.45$\pm$0.12 & 0.52$\pm$1.57 & -2.04$\pm$1.67\\[0.1cm]
    $\log_{10}(\chi_{\rm ray})$& & & 0.14$\pm$0.13 &  2.71$\pm$0.80 & 1.23$\pm$1.23\\[0.1cm]
    $\log_{10}(\chi_{\rm Fe})$& & & -6.05$\pm$0.22  &  -4.40$\pm$0.83  & -6.69$\pm$1.27\\[0.1cm]
    $\log_{10}(\chi_{\rm Mg})$& & & -6.10$\pm$0.27 &  -4.18$\pm$1.06  & -6.58$\pm$1.40\\[0.1cm]
    $\log_{10}(\chi_{\rm Cr})$& & & -8.54$\pm$0.21 &  -6.41$\pm$0.86  & -8.51$\pm$1.36\\[0.1cm]
    $\log_{10}(\chi_{\rm Ti})$& & & -11.72$\pm$0.20 & -8.20$\pm$1.01  & -10.81$\pm$1.47\\[0.1cm]
    $\log_{10}(\chi_{\rm V})$& & & -9.77$\pm$0.18 &  -7.72$\pm$0.82  & -10.13$\pm$1.30\\[0.1cm]
    $\log_{10}(\chi_{\rm Na})$& & & -8.98$\pm$0.24 & -7.10$\pm$0.91  & -9.23$\pm$1.45\\[0.1cm]
    $\log_{10}(\chi_{\rm Ca})$& & & -9.38$\pm$0.22 & -7.22$\pm$0.87  & -10.59$\pm$1.40\\[0.1cm]
    \hline\\[3cm]
    \end{tabular}
    \end{table}

    \begin{figure}
    \centering
        \includegraphics[width=12cm]{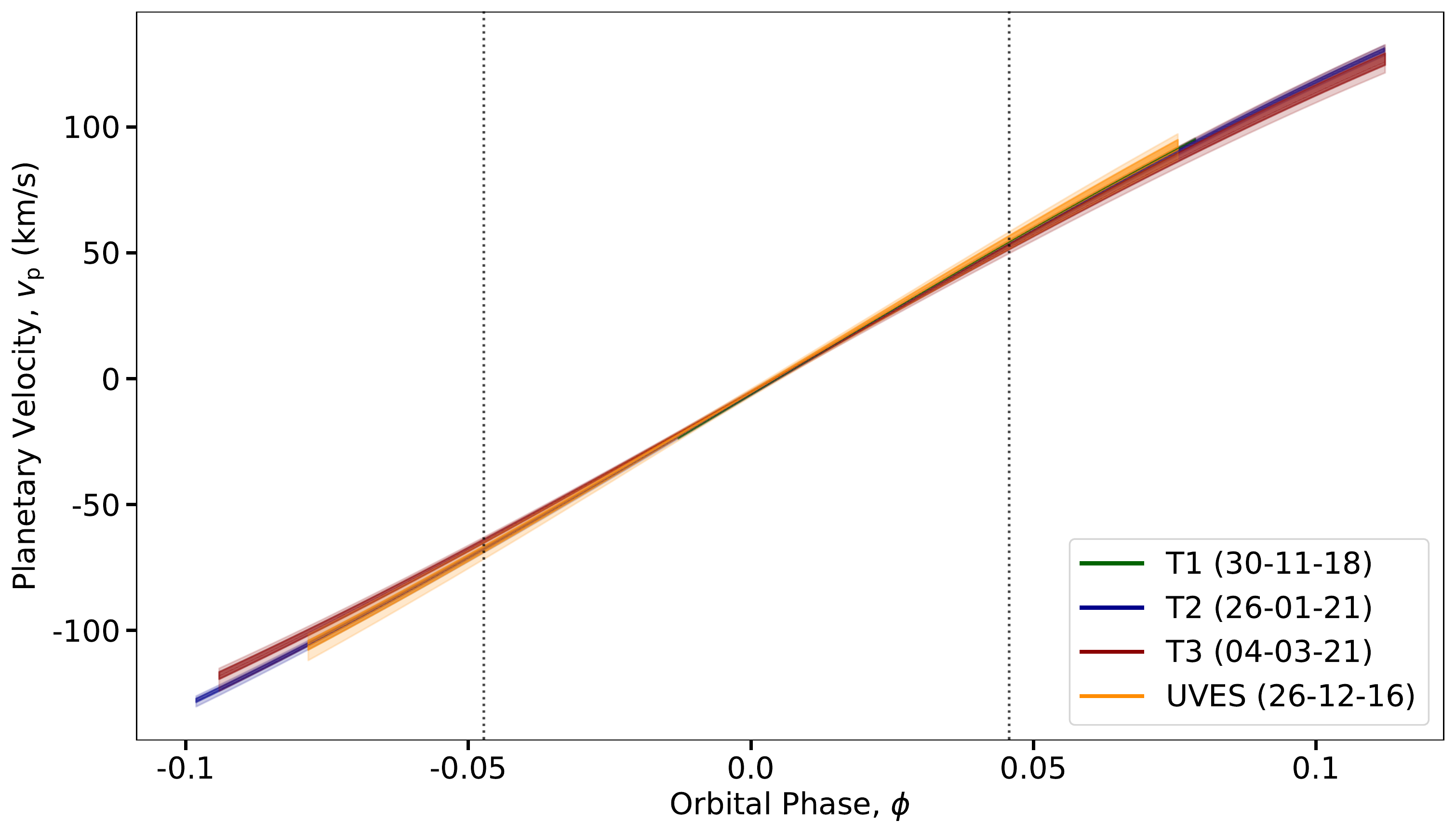}
        \caption{The retrieved planetary orbital velocity from each of our observations, computed from 10,000 random samples of the MCMC, where the solid curves show the median value, and the shaded regions show the 1$\sigma$ contour. The vertical dotted lines indicate the orbital phases of ingress and egress.}
        \label{fig:vp_comp}
    \end{figure}
\end{center}




\label{lastpage}
\end{document}